\documentclass[onecolumn,draftclsnofoot,12pt]{IEEEtran}
\usepackage{acronym}
\usepackage{amsfonts}
\usepackage[dvips]{graphicx}
\usepackage{times}
\usepackage{cite}
\usepackage{amsmath}
\usepackage{array}
\usepackage{amssymb}
\usepackage{stfloats}
\usepackage{slashbox}
\usepackage{graphicx}
\usepackage{footnote}
\usepackage{amsthm}
\usepackage{booktabs}
\usepackage{array}
\usepackage{algorithmic}
\usepackage{algorithm}
\usepackage{subeqnarray}
\usepackage{cases}
\usepackage{threeparttable}
\usepackage{color}
\usepackage{epstopdf}
\usepackage{multirow}
\usepackage{tabularx}
\usepackage{enumerate}
\usepackage{multicol}
\usepackage{subfigure}

\newcommand{\figref}[1]{{Fig.}~\ref{#1}}

\def\bb0{{\mathbb{0}}}

\def\bb{{\mathbf{b}}}

\def\bh{{\mathbf{h}}}

\def\bn{{\mathbf{n}}}

\def\br{{\mathbf{r}}}

\def\bv{{\mathbf{v}}}

\def\by{{\mathbf{y}}}

\def\b0{{\mathbf{0}}}

\def\bA{{\mathbf{A}}}
\def\bB{{\mathbf{B}}}
\def\bC{{\mathbf{C}}}

\def\bH{{\mathbf{H}}}
\def\bI{{\mathbf{I}}}

\def\bQ{{\mathbf{Q}}}

\def\sf0{{\mathsf{0}}}

\def\rm0{{\mathrm{0}}}

\def\Nt{{N_\mathrm{t}}}
\def\Nr{{N_\mathrm{r}}}

\def\Pt{{P_{\mr{t}}}}
\def\Hb{{\mathcal{H}_{\mr{b}}}}

\newcommand{\mb}{\mathbf}
\newcommand{\mr}{\mathrm}
\newcommand{\diag}[1]{\mathrm{diag}\left\{#1\right\}}
\def\j{\mathrm{j}}
\def\Re{\mathrm{Re}}
\def\Im{\mathrm{Im}}
\acrodef{CSI}[CSI]{channel state information}
\acrodef{CSIT}[CSIT]{channel state information at the transmitter}
\acrodef{CSIR}[CSIR]{channel state information at the receiver}
\acrodef{MIMO}[MIMO]{multiple-input multiple-output}
\acrodef{SISO}[SISO]{single-input single-output}
\acrodef{MISO}[MISO]{multiple-input single-output}
\acrodef{SIMO}[SIMO]{single-input multiple-output}
\acrodef{ADCs}[ADCs]{analog-to-digital convertors}
\acrodef{SNR}[SNR]{signal-to-noise ratio}
\acrodef{AWGN}[AWGN]{additive white Gaussian noise}
\acrodef{MRT}[MRT]{maximal ratio transmission}

\IEEEoverridecommandlockouts

\begin{document}
%
%\IEEEoverridecommandlockouts

\title{Limited Feedback in Single and Multi-user MIMO Systems with Finite-Bit ADCs}
	\author{\IEEEauthorblockN{Jianhua~Mo,~\IEEEmembership{Student~Member,~IEEE}, and \IEEEauthorblockN{Robert W. Heath Jr.}, ~\IEEEmembership{Fellow,~IEEE}}\\
	\thanks{Jianhua Mo and Robert W. Heath Jr. are with Wireless Networking and Communications Group, The University of Texas at Austin, Austin, TX 78712, USA (Email: \{jhmo, rheath\}@utexas.edu).This work was supported in part by the National Science Foundation under Grant No. NSF-CCF-1319556 and NSF-CCF-1527079.}
	\thanks{The material in this paper was presented in part at the 2015 and 2016 Asilomar Conference on Signals, Systems and Computers \cite{Mo_Jianhua_Asilomar15,Mo_Jianhua_Asilomar16}.}
}
\maketitle

\begin{abstract}
%Communication systems with low-resolution analog-to-digital-converters (ADCs) can exploit channel state information at the transmitter (CSIT) and receiver. This paper presents initial results on codebook design and performance analysis for limited feedback systems with 1-bit ADCs. Different from the high-resolution case, the absolute phase at the receiver is important to align the phase of the received signals when the received signal is sliced by 1-bit ADCs. A new codebook design for the beamforming case is proposed that separately quantizes the channel direction and the residual phase.

%We analyze limited feedback in systems where a multiple-antenna transmitter sends signals to single-antenna receivers with finite-bit ADCs. If channel state information (CSI) is not available with high resolution at the transmitter and the precoding is not well designed, the inter-user interference is a big decoding challenge for receivers with low-resolution quantization. In this paper, we derive achievable rates with finite-bit ADCs and finite-bit CSI feedback. The performance loss compared to the case with perfect CSI is then analyzed. The results show that the number of bits per feedback should increase linearly with the ADC resolution to restrict the loss.

Communication systems with low-resolution analog-to-digital-converters (ADCs) can exploit channel state information at the transmitter and receiver. This paper presents codebook designs and performance analyses for limited feedback MIMO systems with finite-bit ADCs. 
A point-to-point single-user channel is firstly considered.
When the received signal is sliced by 1-bit ADCs, the absolute phase at the receiver is important to align the phase of the received signals. A new codebook design for beamforming, which separately quantizes the channel direction and the residual phase, is therefore proposed. For the multi-bit case where the optimal transmission method is unknown, suboptimal Gaussian signaling and eigenvector beamforming is assumed to obtain a lower bound of the achievable rate. It is found that to limit the rate loss, more feedback bits are needed in the medium SNR regime than the low and high SNR regimes, which is quite different from the conventional infinite-bit ADC case.
Second, a multi-user system where a multiple-antenna transmitter sends signals to multiple single-antenna receivers with finite-bit ADCs is considered. 
Based on the derived performance loss due to finite-bit ADCs and finite-bit CSI feedback, the number of bits per feedback should increase linearly with the ADC resolution in order to restrict the rate loss.

\end{abstract}


\section{Introduction}

%\jianhuacomment{Consider the QAM transmission with phase offset}
%
%\jianhuacomment{Consider the joint channel estimation and limited feedback MISO channel finite-bit quantization, See Caire\_IT10 paper}
%
%\jianhuacomment{Include channel estimation}

Wide bandwidths and large antenna arrays are two keys to higher achieve transmission rates in future communication systems. At the same time, however, they impose challenges for the hardware design of the receiver, which has to efficiently process signals from multiple antennas (e.g., $\geq 100$ antennas) at a much faster rate (e.g., $\geq 1$ GHz). The analog-to-digital converter (ADC) is a power consumption bottleneck in wideband \ac{MIMO} architectures \cite{Murmann_FTFC13, Murmann_16}. 
%At rates above 100 Mega samples-per-second, ADC power consumption increases quadratically with the sampling frequency \cite{Murmann_FTFC13, Murmann_16}.
%
%As communication systems are being designed with ever wider bandwidths and ever more antennas, the device that mediates between the worlds of analog and digital -- the analog-to-digital converter (ADC) -- is being rapidly pushed to its limits.  At rates above 100 Mega samples per second, ADC power consumption increases quadratically with sampling frequency \cite{Murmann_FTFC13}. While the absolute energy per conversion step has decreased, the rate of energy increase has seen little improvement in the past decade \cite{Murmann_16}. One approach to overcoming this challenge is to live with low-resolution ADCs.
%Such approaches can be implemented using today's technology without requiring serendipity in circuits or semiconductors.
%
The use of few- and especially 1-bit ADCs is one possible approach to overcoming this bottleneck. Low resolution ADCs have been widely explored in millimeter wave \ac{MIMO} systems \cite{Singh_TCOM09, Mezghani_ISIT07, Dabeer_ICC10, Mezghani_WSA10, Mezghani_ISIT12, Bai_Qing_ETT15, Mo_Jianhua_TSP15, Mo_Jianhua_TWC17, Mo_Jianhua_arxiv16b, Wen_Chao-Kai_TSP16,Rusu_Asilomar15, Rini_arxiv17} and massive \ac{MIMO} systems \cite{Choi_TCOM16, Mollen_TWC17, Studer_TCOM16, Jacobsson_ICC15, Fan_Li_CL15, Wang_Shengchu_TWC15}. Prior work has shown that low resolution ADCs are practical for wireless communications. It was found that there is negligible SNR and rate loss (for example, less than $2$ dB for 1-bit quantization) at low SNR compared to infinite-bit ADCs \cite{Singh_TCOM09,Mo_Jianhua_TSP15}. It is also possible to estimate the channel (IID Rayleigh fading or correlated, narrowband or broadband) \cite{Dabeer_ICC10, Mezghani_WSA10, Wen_Chao-Kai_TSP16, Studer_TCOM16,Mo_Jianhua_arxiv16b,Rusu_Asilomar15,Mollen_TWC17}, and detect symbols (QPSK or higher-order QAM) with coarse quantization \cite{Wen_Chao-Kai_TSP16,Wang_Shengchu_TWC15,Choi_TCOM16}.

At present, the capacity of the quantized \ac{MIMO} channel with \ac{CSIT} is generally unknown, except for the simple \ac{MISO} channel and some special cases, such as in the low or high SNR regime \cite{Dabeer_SPAWC06, Mezghani_ISIT07, Singh_TCOM09, Mo_Jianhua_TSP15, Rini_arxiv17}. %Transmitting independent QPSK signals \cite{Mezghani_ISIT07} or Gaussian signals \cite{Mezghani_ISIT12, Bai_Qing_ETT15} from each antenna nearly achieves the capacity at low SNR, but is far from the optimal at high SNR.
Our previous work \cite{Mo_Jianhua_TSP15} shows that maximum ratio transmission (MRT) achieves the capacity of quantized MISO channels. 
It was also suggested in \cite{Mo_Jianhua_TSP15} that channel-inverse precoding (or called zero-forcing precoding), which eliminates the inter-stream interference before the low-resolution quantization, provides a substantial performance improvement compared with the no-precoding case. CSIT, though, is required for both the maximum ratio transmission and channel-inverse precoding. In multi-user systems, if CSIT is not accurate enough and thus precoding is not well designed, the inter-user interference is a decoding challenge for receivers with low-resolution ADCs \cite{Choi_TCOM16}. Therefore, CSIT is preferred in quantized single and multi-user MIMO channels.

Despite the potential gain from transmitter precoding, there is little work on limited feedback with low-resolution ADCs, besides our initial results in \cite{Mo_Jianhua_Asilomar15,Mo_Jianhua_Asilomar16}. The results on limited feedback with infinite-resolution ADCs, e.g., \cite{Love_IT03, Jindal_IT06, Love_JSAC08}, cannot be directly extended to low-resolution ADCs since codebook requirements and the achievable rate expressions are different. For example, in MISO limited feedback beamforming with 1-bit ADCs, the optimum beamformer is phase invariant, meaning that equivalent performance is achieved by $\bv$ and $\bv e^{j \theta}$. The optimum beamformer, though, is the matched filter, and is not phase invariant. The reason is that phase at the receiver is important for detecting QPSK signals using 1-bit ADCs. A key function of \ac{CSIT} is to align the phase of the received signals such that the real and imaginary parts are quantized independently. As a result, phase-invariant Grassmannian beamforming codebooks \cite{Love_IT03} are no longer appropriate. 
The achievable rate for channels with low resolution ADCs is different from those with infinite resolution ADCs. For example, at high SNR, the achievable rate of the channel is limited by the ADC resolution. 
%Therefore, scaling laws with conventional infinite-bit ADCs, which says that which say the bits per feedback must increase linearly with the SNR (in decibels) to limit the rate loss in multi-user MISO systems, 
%cannot be applied here. 
Based on our analyses, for the single-user channel, more feedback bits are actually needed in the medium SNR regime, while in the low and high SNR regimes, accurate \ac{CSIT} is not needed to reduce the rate loss.

In this paper, we develop limited feedback methods for multiple-antenna systems with few-bit ADCs.
Our approach leverages recent work \cite{Dabeer_ICC10, Mezghani_WSA10, Wen_Chao-Kai_TSP16,Studer_TCOM16,Mo_Jianhua_arxiv16b,Rusu_Asilomar15,Mollen_TWC17} that shows that it is possible to estimate narrowband and broadband MIMO channels even with 1-bit ADCs at the receiver.
%using idea from 1-bit compressive sensing \cite{Boufounos_CISS08, Jacques_IT13}.
Given a perfect estimate of the channel, we propose limited feedback methods based on the explicit and implicit approaches of dealing with the ADC quantization impairment for flat-fading channels.
Our work provides a path to making the assumption of CSIT in multiple-antenna systems with few-bit ADCs more realistic.

We use two different analytical approaches for computing the achievable rate. The first approach accounts for nonlinear ADC operation in the system optimization. The signal design at the transmitter, and the ADC design (including the thresholds and reconstruction points) at the receiver are specifically optimized. For example, the optimal input signaling is discrete, with at most $K+1$ points where $K$ is the number of possible quantization outputs at the receiver \cite{Mo_Jianhua_TSP15,Singh_TCOM09}. This approach can provide the exact expressions of the system performance, for example, the channel capacity, bit error rate, etc.
This applies well for the case of 1-bit ADCs since the optimal signaling is known to be QPSK, but not the case of multi-bit ADC where the optimal signaling and ADC design is unknown, though lower bounds of the achievable rates can be obtained \cite{Rini_arxiv17,Mo_Jianhua_TWC17} by using the suboptimal QAM signaling and uniform quantization. Iterative numerical method can be used to give a suboptimal solution but the optimality is not guaranteed \cite{Singh_TCOM09, Mo_Jianhua_TSP15}.
Another implicit and approximate approach models the quantization as an additional impairment to the system. In the second approach, the quantization process is linearized by the MMSE estimator and the quantization noise is introduced to model the signal distortion. With this option, the first and second order statistics of the signal and its quantized version are preserved. Usually complex Gaussian signaling is assumed at the transmitters which is the same as in conventional full-resolution systems. The quantization noise is also assumed to be the worst-case Gaussian distributed to provide an lower bound of the system performance. The lower bound is generally tight enough for multi-bit ADC in the low SNR regime \cite{Mezghani_ISIT12,Bai_Qing_ETT15}. For these two approaches, we develop different limited feedback methods.

The contributions of this paper are summarized as follows.
\begin{enumerate}
	\item We analyze single-user \ac{SISO} and \ac{MISO} channels with \emph{1-bit} ADCs where the transmitter sends capacity-achieving QPSK symbols. Our proposed codebook design for the \ac{MISO} beamforming case separately quantizes the channel direction and the residual phase to incorporate the phase sensitivity of QPSK symbols. Bounds of the power and rate loss with respect to the number of feedback bits
	are derived.
	%This design, however, cannot be extended to the channel with more than 1-bit ADCs because the optimal signaling in this case is unknown.
	\item We analyze single-user channels with multi-bit ADCs by assuming that the transmitter adopts suboptimal complex Gaussian signaling. Since complex Gaussian signaling is circularly symmetric, a single codebook quantizing the channel direction is enough.
	%We derive bounds on the achievable rate with finite-bit ADCs and infinite-bit feedback. 
	The rate and power losses incurred by the finite rate feedback compared to perfect CSIT is also found. 
	%The results bridge the gap between the case of infinite-bit ADC \cite{Jindal_IT06} and 1-bit ADC.
	\item We analyze limited feedback in multi-user systems where a multiple-antenna transmitter sends signals to multiple single-antenna receivers with finite-bit ADCs.  We derive achievable rates with finite-bit ADCs and finite-rate CSI feedback. The performance loss compared to the case with perfect CSI is then analyzed. The results show that the number of bits per feedback should be increased linearly with the ADC resolution to restrict the rate loss.
\end{enumerate}

%The material in this paper is present partly in \cite{Mo_Jianhua_Asilomar15,Mo_Jianhua_Asilomar16} where SISO and MISO cases are considered. In this paper, we include the results of MIMO channel where the derivation is much more complicated.

The results on SISO and MISO channels were presented partly in \cite{Mo_Jianhua_Asilomar15, Mo_Jianhua_Asilomar16}. In this paper, we extend the previous results and includes our new analyses for MIMO channel. Different from the single receiver antenna case, in the MIMO channel there are multiple correlated received signals. The low-resolution quantization will impact this correlation and brings challenges in the rate analyses. We investigate this effect in our design of limited feedback method and transmitter beamforming. 

\emph{Organization}: In Section II, the system model and ADC setup is shown. In Section III and IV, the single-user and multi-user systems are investigated. Simulation results are provided in Section V to verify our analyses. The paper is summarized in Section VI.

	\emph{Notation}: $a$ is a scalar, $\mb{a}$ is a vector and $\mb{A}$ is a matrix.
$\angle x$ represents the phase of a complex number $x$. $\Re(x)$ and $\Im(x)$ denote the real and imaginary part of $x$, respectively.
%$\mb{x}_{i:j}$ is the vector consisting of \{$x_k$, $i \leq k \leq j$\}.
$\mr{tr}(\mb{A})$, $\mb{A}^T$, $\mb{A}^*$ and $||\mb{A}||_F$ represent the trace, transpose, conjugate transpose and Frobenius norm of a matrix $\mb{A}$, while $\diag{\bA}$ represents a diagonal matrix by keeping only the diagonal elements of $\bA$.

\section{System Model}
In this paper, we consider single-user MISO, MIMO and multiple-user MISO systems. The transmitter is equipped with $\Nt$ antennas, while each receiver has $\Nr$ antennas with finite-bit ADCs. There are $2\Nr$ $b$-bit ADCs that separately quantize the real and imaginary part of the received signal. We assume that uniform quantization is applied since it is easier for implementation and achieves only slightly worse performance than the non-uniform case\cite{Max_IRE60}.
\begin{table*}
	\centering
	\caption{The optimum uniform quantizer for a Gaussian zero-mean unit-variance signal \cite{Max_IRE60} }
	\label{tab:Eta_b}
	\begin{tabular}{|c|c|c|c|c|c|c|c|c|}
		\hline
		Resolution $b$  & 1-bit  & 2-bit  & 3-bit & 4-bit & 5-bit & 6-bit & 7-bit & 8-bit\\
		\hline
		NMSE $\eta_b$ & $\frac{\pi-2}{\pi} \left(\approx 0.3634 \right)$ & 0.1175 & 0.03454 & 0.009497 & 0.002499 & 0.0006642 & 0.0001660 & 0.00004151\\
		\hline
		%		SQR $\frac{1}{\eta_b}$ (dB) & 4.40 & 9.30 &  14.62 &  20.22 &  26.02 &  31.78  & 37.80 &  43.82\\
		%		\hline
		Stepsize $\Delta_b$  & $\sqrt{\frac{8}{\pi}}$ ($\approx$1.5958) & 0.9957 & 0.586 & 0.3352 & 0.1881 & 0.1041 & 0.0569 & 0.0308 \\
		\hline
		$10 \log_{10} \left(1 -\eta_b \right)$ & $10 \log_{10} \frac{2}{\pi}$ ($\approx$-1.9613)  & -0.5429  & -0.1527  & -0.0414  & -0.0109  & -0.0029  & -0.0007 &  -0.0002 \\
		\hline
		$\log_{2} \left(\frac{1}{\eta_b}\right) $ & $\log_{2} \frac{\pi}{\pi-2}$ ($\approx$ 1.46) & 3.09 &  4.86 &  6.72 &  8.64 &  10.56  & 12.56 &  14.56\\
		\hline
	\end{tabular}
\end{table*}
%For multi-bit quantization, we assume in this paper that uniform quantization is used.
For a complex-valued scalar $x$, we say $y=\mathcal{Q}(x)$ if
\begin{align}\label{eq:few-bit_ADC}
	y &=  \mr{sign} \left( \Re(x) \right)  \left( \min \left( \left\lceil \frac{|\Re(x)|}{\Delta_{\mr{Re}}} \right \rceil, 2^{b-1} \right)  - \frac{1}{2} \right) \Delta_{\mr{Re}} \nonumber \\
	&\quad + \j \; \mr{sign} \left( \Im(x) \right)  \left( \min \left( \left\lceil \frac{|\Im(x)|}{\Delta_{\mr{Im}}} \right \rceil, 2^{b-1} \right)  - \frac{1}{2} \right) \Delta_{\mr{Im}}.
\end{align}
%	\begin{multline}
%	y =  \mr{sign} \left( \Re(x) \right)  \left( \min \left( \left\lceil \frac{|\Re(x)|}{\Delta} \right \rceil, 2^{b-1} \right)  - \frac{1}{2} \right) \Delta \\
%	+ \j \; \mr{sign} \left( \Im(x) \right)  \left( \min \left( \left\lceil \frac{|\Im(x)|}{\Delta} \right \rceil, 2^{b-1} \right)  - \frac{1}{2} \right) \Delta.
%	\end{multline}
%	For a real-valued scalar $x$, $y=\mathcal{Q}(x)$ if
%	\begin{align}
%	y = \mr{sign} \left(x \right)  \left( \min \left( \left\lceil \frac{|x|}{\Delta} \right \rceil, 2^{b-1} \right)  - \frac{1}{2} \right) \Delta.
%	\end{align}
where $\Delta_{\mr{Re}} = \Delta_b  \left( \mathbb{E} \left[ |\Re(x)|^2 \right] \right)^{\frac{1}{2}}$ and $\Delta_{\mr{Im}} = \Delta_b  \left( \mathbb{E} \left[ |\Im(x)|^2 \right] \right)^{\frac{1}{2}}$. The average power of the received signal, i.e., $\mathbb{E} \left[ |\Re(x)|^2 \right]$ and $\mathbb{E} \left[ |\Im(x)|^2 \right]$, can be detected by the analog circuits before the ADCs, for example, automatic gain control (AGC). 
For the special case of 1-bit quantization,
\begin{align}\label{eq:1-bit_ADC}
	y &=  \mr{sign} \left( \Re(x) \right) \sqrt{\frac{2}{\pi}} \left( \mathbb{E} \left[ |\Re(x)|^2 \right] \right)^{\frac{1}{2}} + \j \; \mr{sign} \left( \Im(x) \right) \sqrt{\frac{2}{\pi}} \left( \mathbb{E} \left[ |\Im(x)|^2 \right] \right)^{\frac{1}{2}}.
\end{align}
For a circularly symmetric signal, which is considered in this paper, $\mathbb{E} \left[ |\Re(x)|^2 \right]= \mathbb{E} \left[ |\Im(x)|^2 \right] = \frac{1}{2} \mathbb{E} \left[ |x|^2 \right]$ and therefore $\Delta_{\mr{Re}} = \Delta_{\mr{Im}}$.
The quantization stepsize $\Delta_b$ is chosen to minimized the mean squared error for a unit-norm Gaussian input signal (see \cite{Max_IRE60}).
These values of $\Delta_b$ are given in Table \ref{tab:Eta_b} assuming the input signal has unit-power. The reconstruction points, as shown in \eqref{eq:few-bit_ADC} are the middle points between two adjacent quantization thresholds.
The normalized mean squared error (NMSE), denoted as $\eta_b\triangleq \frac{\mathbb{E} \left[ \left| \mathcal{Q}(x) -  x \right|^2 \right]}{\mathbb{E} \left[ \left| x \right| ^2 \right]}$ is also listed. 

Throughout this paper, we assume that the channel follows IID Rayleigh fading. The extension to correlated channel model is an interesting topic for future work.
We also assume the receiver has perfect channel state information. This is justified by prior work on channel estimation with low resolution ADCs, for example \cite{Dabeer_ICC10, Mezghani_WSA10, Wen_Chao-Kai_TSP16, Studer_TCOM16,Mo_Jianhua_arxiv16b}. Furthermore, the feedback is assumed to be delay and error free, as is typical in limited feedback problems. Adding realism to the feedback channel is an interesting topic for future work.

%\begin{figure}
%    \begin{subfigure}{.49\textwidth}
%        \centering
%        \includegraphics[width=\linewidth]{Figures/SU_MISO.eps}
%        \caption{Single-user MISO system} \label{fig:SU_MISO}
%    \end{subfigure}\hfill
%    \begin{subfigure}{.49\textwidth}
%        \centering
%        \includegraphics[width=0.95\linewidth]{Figures/MU_MISO.eps}
%        \caption{Multi-user MISO system} \label{fig:MU_MISO}
%    \end{subfigure}
%    \caption{MISO systems with finite-bit quantization and limited feedback. At each receiver, there are two $b$-bit ADCs. There is also a low-rate feedback path from each receiver to the transmitter.} \label{fig:MISO}
%\end{figure}

%\begin{figure}[t]
%	\centering
%	\subfigure[$\Nt=4$]{
%		\includegraphics[width=0.7\columnwidth]{Figures/MISO_Capacity_Nt_4.eps}
%		\label{fig:MISO_Capacity_Nt_4}
%	}
%	\subfigure[$\Nt=16$]{
%		\includegraphics[width=0.7\columnwidth]{Figures/MISO_Capacity_Nt_16.eps}
%		\label{fig:MISO_Capacity_Nt_16}
%	}
%	\caption{The achievable rates of a MISO system with CSIT, no CSIT and limited feedback when the ADC resolution is 1-bit.}
%	\label{fig:MISO_Capacity_1bit_ADC}
%\end{figure}

%For the case of single-user MISO channel, the optimal choice 

\section{Single-user Channel with Finite-bit ADCs and Limited Feedback} \label{sec:SU_MISO}

\begin{figure}[t]
	\begin{centering}
		\includegraphics[width=0.8\columnwidth]{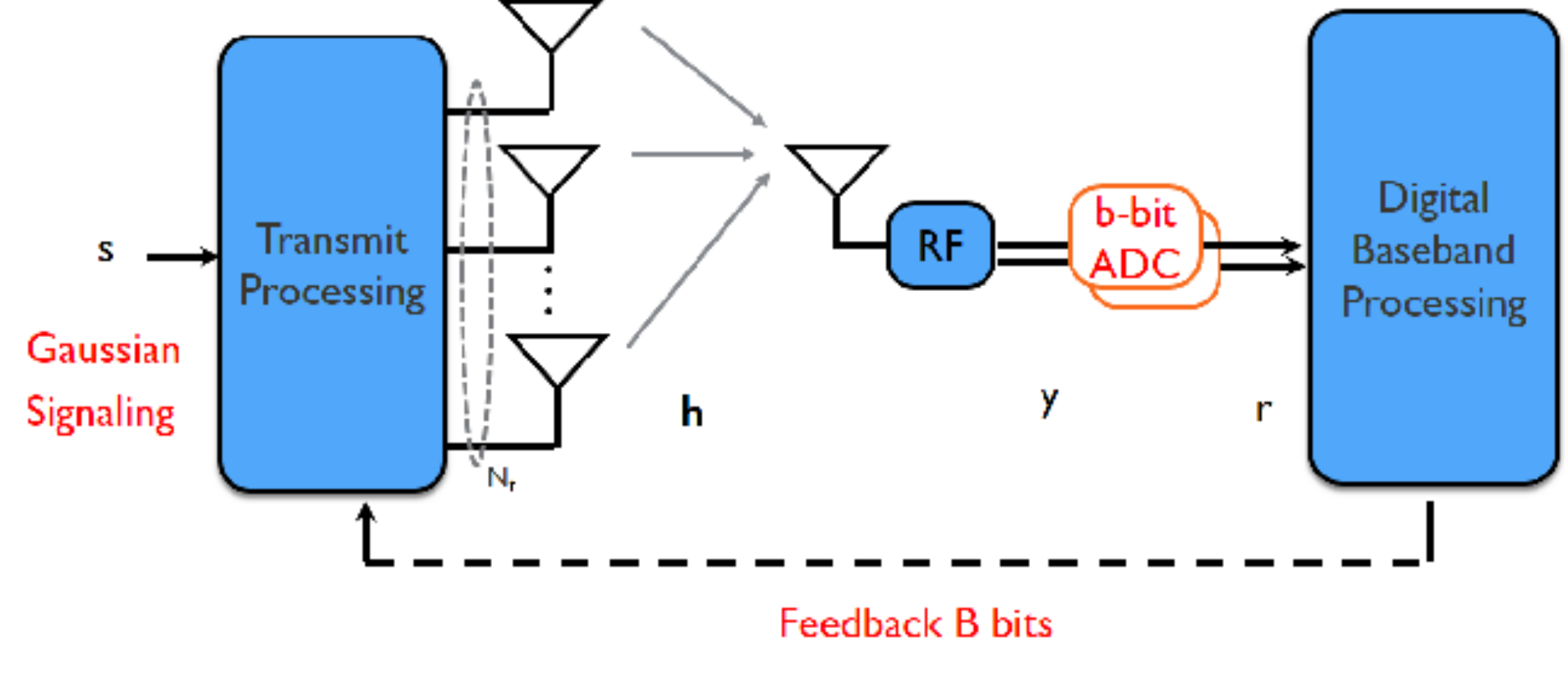}
		\vspace{-0.5cm}
		\centering
		\caption{A MISO system with finite-bit quantization and limited feedback. At the receiver side, there are two $b$-bit ADCs. There is also a low-rate feedback path from the receiver to the transmitter. Note that there is no limitation on the structure of the transmitter.}\label{fig:SU_MISO}
	\end{centering}
	\vspace{-0.3cm}
\end{figure}

In this section, we consider a single-user MISO system with finite-bit quantization, as shown in Fig. \ref{fig:SU_MISO}.
\subsection{MISO Channel with Finite-Bit Quantization and Limited Feedback}

%	\begin{figure}
%		\centering
%		\subfigure[]{
%			\includegraphics[width=0.95\columnwidth]{Figures/SU_MISO.eps}
%			\label{fig:Spatial_Domain_Channel}
%		}
%		\subfigure[]{
%			\includegraphics[width=0.95\columnwidth]{Figures/MU_MISO.eps}
%			\label{fig:Angular_Domain_Channel}
%		}
%		\caption{The figure shows an example of $64 \times 64$ narrowband ULA MIMO channel. The magnitude of each element of the aperture domain channel $\bH$ is shown in (a) where the signal is obviously not sparse. The angular domain channel $\bX$ is shown in (b) where there are only two groups of entries with large magnitudes, each of which corresponds to one cluster of the wireless channel.}\label{fig:Beam_sparse}
%	\end{figure}

%There are $\Nt$ antennas at the transmitter and single antenna at the receiver.
Assuming perfect synchronization and a narrowband channel, the baseband received signal in this MISO system is
\begin{equation}
y = \mb{h}^*\mb{v} s + n,
\end{equation}
where $\mb{h} \in \mathbb{C}^{\Nt\times 1}$ is the channel vector, $\mb{v}\in \mathbb{C}^{\Nt \times 1} (\| \bv\|=1)$ is the beamforming vector, $s$ is the Gaussian distributed symbol sent by the transmitter, $y\in \mathbb{C}$ is the received signal before quantization, and $n \sim \mathcal{CN}(0, \sigma_n^2)$ is the circularly symmetric complex Gaussian noise. The signal variance is $\mathbb{E}[|s|^2] = \Pt$.

The received signal after finite-bit quantization is
\begin{equation}
r = \mathcal{Q}\left(y\right) = \mathcal{Q} \left(\mb{h}^*\mb{v} s + n\right).
\end{equation}
%where $\mathcal{Q}(\cdot)$ is the finite-bit quantization function applied separately to the real and imaginary parts.
%In our system, there are two finite-bit resolution quantizers that separately quantize the real and imaginary part of the received signal.
%We assume that uniform quantization is applied since it is easier for implementation achieves only slightly worse performance than non-uniform case\cite{Max_IRE60}.

\subsubsection{Single-Bit Quantization}

We first consider the SISO channel as a special case. Since  $\Nt=\Nr=1$, the received signal is
\begin{align}
r = \mr{sgn}(y) = \mr{sgn}(h x + n).
\end{align}
where $\mr{sgn}(y)$ provides the sign of the real and imaginary parts of $y$. We eliminate the amplitude in \eqref{eq:1-bit_ADC} for simplicity as it does not affect the capacity analyses.
Denote the phase of $h$ as $\angle h$. As shown in our previous work \cite[Lemma 1]{Mo_Jianhua_TSP15}, the capacity-achieving input is equally likely rotated QPSK symbols,
\begin{equation}
\mr{Pr}\left[x = \sqrt{\Pt} e^{\j\left( \frac{k \pi}{2} + \frac{\pi}{4} - \angle h \right)}\right] = \frac{1}{4}, \; \mr{for} \; k=0, 1, 2 \;\mr{and}\; 3.
\end{equation}
The term $-\angle h$ in the transmitted signals is introduced to pre-cancel the phase rotation of the channel such that the receiver will observe a regular QPSK signal.

If $h$ is unknown at the transmitter, then only the phase needs to be quantized and fed back to the transmitter. Since the QPSK constellation is unchanged for a 90-degree rotation, only $\mr{mod} \left(\angle h, \frac{\pi}{2}\right)$ instead of $\angle h$ needs to be fed back. Now assume $B$ bits are used to uniformly quantize the region $[0, \frac{\pi}{2}]$. Uniform quantization is reasonable since for most statistical channel models the phase of the SISO channel is uniformly distributed. The codebook is then $\Psi=\{\psi_i = \frac{i\pi}{2^{B+1}} + \frac{\pi}{2^{B+2}}, 0 \leq i \leq 2^B-1 \}$. For instance, $\Psi=\{\frac{\pi}{8}, \frac{3\pi}{8}\}$ if $B=1$.
The receiver sends the index $i$ of $\widehat{\psi}$ to the transmitter such that
\begin{eqnarray}
\widehat{\psi} = \underset{\psi_m \in \Psi}{\arg \min} \left|\mathrm{mod} \left(\angle h, \frac{\pi}{2} \right) - \psi_m \right|.
\end{eqnarray}
Based on the feedback index $i$, the transmitter sends rotated QPSK signals with uniform probabilities, i.e.,
\begin{equation}
\mr{Pr}\left[x = \sqrt{\Pt} e^{\j\left( \frac{k \pi}{2} + \frac{\pi}{4} - \widehat{\psi} \right)}\right] = \frac{1}{4}, \; \mr{for} \; k=0, 1, 2 \;\mr{and}\; 3.
\end{equation}
The received signal after quantization is
\begin{eqnarray}
r = \mr{sgn} \left( |h| \sqrt{\Pt} e^{\j\left( \frac{k \pi}{2} + \frac{\pi}{4} - \widehat{\psi} + \angle h \right)} + n \right).
\end{eqnarray}

The channel has four possible inputs and four possible outputs. This is a discrete-input discrete-output channel. Denote the achievable rate with $b$-bit ADC and $B$-bit feedback as $R(b,B)$. Throughout this paper, `$b=\infty$' represents the case of full-precision ADCs, while `$B=\infty$' represents the case of perfect CSIT.
Therefore the achievable rate is
\begin{IEEEeqnarray}{rCl}
R_{\mr{SISO}}(1,B)
&=& 2 - \Hb \left(  Q\left( \sqrt{2 \gamma \left|h\right|^2 \sin^2 \left( \frac{\pi}{4}- \theta \right)}\right) \right) - \Hb \left(  Q\left(\sqrt{2 \gamma  \left|h\right|^2 \cos^2 \left(\frac{\pi}{4} - \theta \right)}\right) \right) \nonumber \\
&=& 2 - \Hb \left(  Q\left(\sqrt{\gamma  \left|h\right|^2 \left(1 - \sin 2 \theta \right)}\right) \right) - \Hb \left(  Q\left( \sqrt{ \gamma \left|h\right|^2 \left(1 + \sin 2 \theta \right)}\right) \right),
\end{IEEEeqnarray}
where $\gamma \triangleq \frac{\Pt}{\sigma_n^2}$, $\theta \triangleq \widehat{\psi} - \mathrm{mod} \left(\angle h, \frac{\pi}{2} \right)$ is the quantization error, $\Hb(p) \triangleq -p \log_2 p -(1-p) \log_2 (1-p)$ is the binary entropy function, and $Q(\cdot)$ is the tail probability of the standard normal distribution.

\begin{figure}[t]
	\begin{centering}
		\includegraphics[width = 0.8\columnwidth]{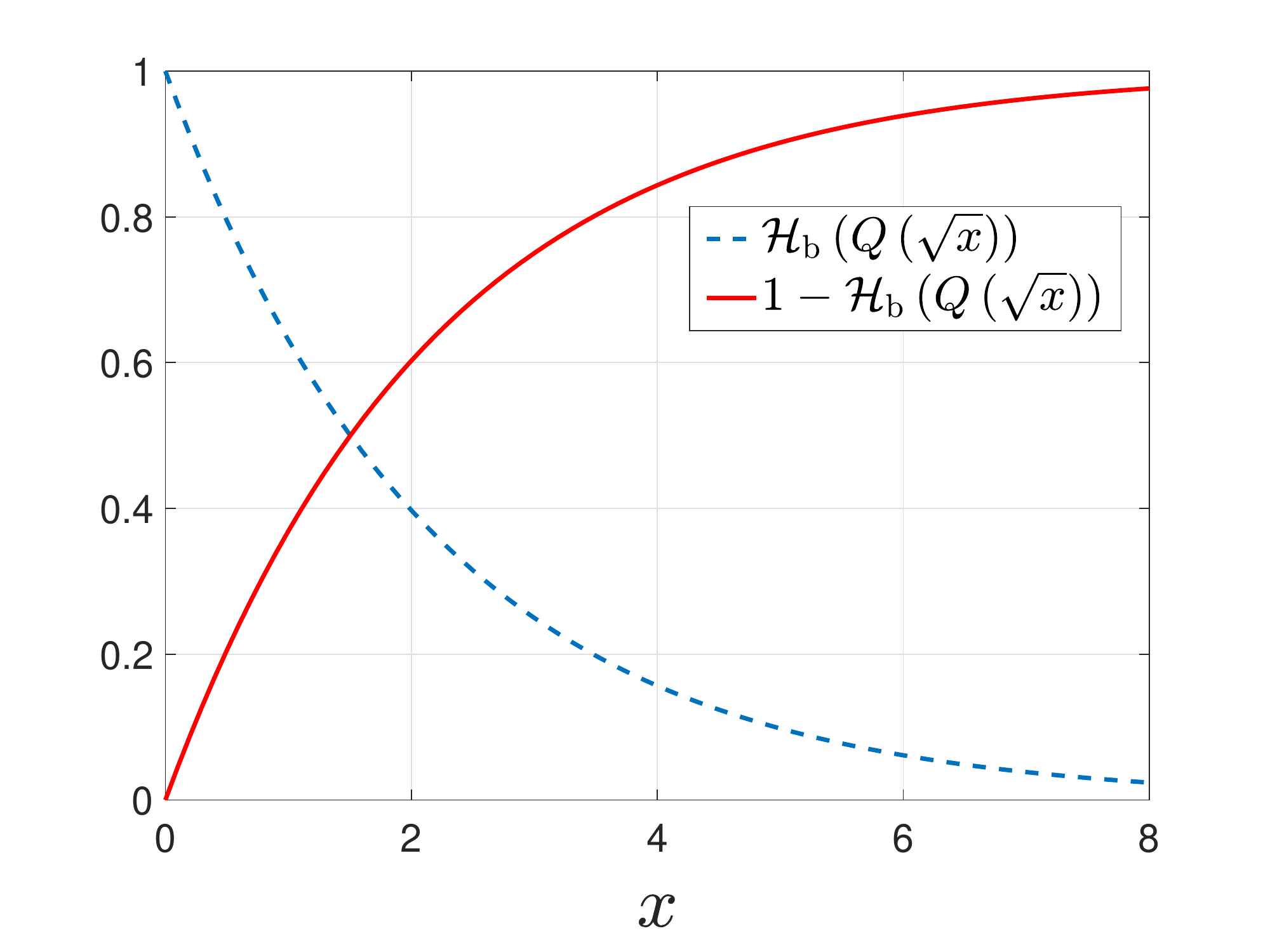}
		\vspace{-0.1cm}
		\centering
		\caption{The figure shows $\Hb \left(Q \left( \sqrt{x}\right) \right)$ and $1-\Hb \left(Q \left( \sqrt{x}\right) \right)$ versus $x$. It is seen that $\Hb \left(Q \left( \sqrt{x}\right) \right)$ is a decreasing and convex function of $x$.
			%In addition, $\Hb \left(Q \left( \sqrt{5}\right) \right) \approx 0.1$.
		}\label{fig:H_b_Q_sqrt_x}
	\end{centering}
	\vspace{-0.3cm}
\end{figure}

In Fig. \ref{fig:H_b_Q_sqrt_x}, we plot the function $\Hb \left(Q \left( \sqrt{x}\right) \right)$.
Since $\Hb \left(Q \left( \sqrt{x}\right) \right)$ is decreasing with $x$, it follows that
\begin{eqnarray} \label{eq:Capacity_SISO_fb_lb}
R_{\mr{SISO}}(1,B) \geq 2 \left(1 - \Hb \left(  Q\left(\sqrt{\gamma  \left|h\right|^2 \left(1 - \sin 2 \left|\theta \right| \right)}\right) \right) \right).
\end{eqnarray}
The channel capacity with perfect \ac{CSIT} is \cite[Lemma 1]{Mo_Jianhua_TSP15}
\begin{equation} \label{eq:Capacity_SISO}
R_{\mr{SISO}}(1, \infty) = 2\left(1 - \Hb \left(  Q\left(\sqrt{\gamma \left|h\right|^2} \right) \right)\right).
\end{equation}

Comparing \eqref{eq:Capacity_SISO_fb_lb} and \eqref{eq:Capacity_SISO}, the power loss factor is $\frac{1}{1- \sin 2 \left| \theta \right|} $. We want to minimize the power loss, or equivalently maximize the term $1 - \sin 2 \left|\theta \right|$.
Since the quantization error $\theta \in \left[ -\frac{\pi}{2^{B+2}}, \frac{\pi}{2^{B+2}} \right]$,
it follows that ${1- \sin 2 \left| \theta \right|} > {1- \sin \frac{\pi}{2^{B+1}}}$. When $B=1$, $1- \sin 2 |\theta| \geq 1 - \frac{1}{\sqrt{2}}$, which means that there is at most $5.33$ $\mr{dB}$ power loss. In addition, as $\theta$ is uniformly distributed, the average power loss is
\begin{eqnarray}
\mathbb{E}_{\theta} \left[ 1 - \sin 2 |\theta| \right]
&=& \frac{1}{\frac{2 \pi}{2^{B+2}}} \int_{-\frac{\pi}{2^{B+2}}}^{\frac{\pi}{2^{B+2}}} \left(1- \sin 2 |\theta| \right)  \mr{d} \theta \\
&=& 1 - \frac{\sin^2 \left(\frac{\pi}{2^{B+2}}\right)}{\frac{\pi}{2^{B+2}}} \label{eq:phase_quan_bound_exact}\\
&\stackrel{(a)}{>}& 1 - \frac{\pi}{2^{B+2}} \\
&>& 1 - 2^{-B} \label{eq:phase_quan_bound_loose} ,
\end{eqnarray}
where $(a)$ follows from $\sin x < x$ for $0< x < \frac{\pi}{2}$.
Therefore, the average power loss is at most $3$ $\mr{dB}$ with only one bit feedback\footnote{A tighter bound is $2$ $\mr{dB}$ by evaluating \eqref{eq:phase_quan_bound_exact} with $B=1$.}.
In the simulation, we will show that with only one bit feedback, the performance is close to that with perfect \ac{CSIT}.

Similar to the MISO system with infinite-resolution ADCs, random vector quantization (RVQ), which is amenable to analysis \cite{Jindal_IT06, Au-Yeung_TWC07}, is assumed to quantize the direction of channel $\bh$. We assume that $B_1$ out of the total $B$ bits are used to convey the channel direction information. The codebook is $\mathcal{W} = \left\{\bv_0, \bv_1, \cdots, \bv_{2^{B_1}-1} \right\}$ where each of the quantization vectors is independently chosen from the isotropic distribution on the Grassmannian manifold $\mathcal{G}(\Nt, 1)$ \cite{Love_IT03}. The receiver sends back the index of $\bv$ maximizing $|\bh^* \bv|$.

Besides the channel direction information, the remaining $B_2 = B -B_1$ bits are used to quantized the phase of the equivalent channel, i.e., $\angle \left(\bh^* \bv \right)$ (denoted as \emph{residual phase} afterwards).
The second codebook quantizing the residual phase is $\Psi=\{\psi_i = \frac{i\pi}{2^{B+1}} + \frac{\pi}{2^{B+2}}, 0 \leq i \leq 2^{B_2}-1 \}$.
The receiver feeds back the index $i$ of $\widehat{\psi}$ such that
\begin{eqnarray}
\widehat{\psi} = \underset{\psi_m}{\arg \min} \left| \mathrm{mod} \left(\angle \left( \bh^* \bv \right), \frac{\pi}{2} \right) - \psi_m \right|.
\end{eqnarray}
The transmitter adopts matched filter beamforming and QPSK signaling based on the feedback bits, i.e.,
\begin{equation}
\mr{Pr}\left[\mb{x} =  \sqrt{\Pt} \bv e^{\j \left( \frac{k \pi}{2} + \frac{\pi}{4} - \widehat{\psi} \right)}\right] = \frac{1}{4}, \; \mr{for} \; k=0, 1, 2 \; \mr{and} \; 3.
\end{equation}
The received signal after quantization is
\begin{eqnarray}
r = \mr{sgn} \left( \sqrt{\Pt} \bh^*\bv e^{\j\left( \frac{k \pi}{2} + \frac{\pi}{4} - \widehat{\psi} \right)} + n \right).
\end{eqnarray}

Similar to the SISO case, this channel is also a discrete-input discrete-output channel. The achievable rate is
\begin{IEEEeqnarray} {rCl}
& &R_{\mr{MISO}}(1,B) \\
&=& 2 - \Hb \left(  Q\left(\sqrt{2 \gamma |\bh^* \bv |^2 \sin^2 \left( \frac{\pi}{4}- \theta \right)}\right) \right) - \Hb \left(  Q\left( \sqrt{2 \gamma |\bh^* \bv |^2 \cos^2 \left( \frac{\pi}{4} - \theta \right)}\right) \right) \\
&=& 2 - \Hb \left(  Q\left(\sqrt{\gamma \|\bh \|^2 \cos^2 \beta \left(1 - \sin 2 \theta \right)}\right) \right) - \Hb \left(  Q\left(\sqrt{ \gamma \|\bh \|^2 \cos^2 \beta \left(1 + \sin 2 \theta \right)}\right) \right) ,
\end{IEEEeqnarray}
where $\theta \triangleq \widehat{\psi} - \mr{mod} \left(\angle \left( \bh^* \bv \right), \frac{\pi}{2}\right)$ and $\cos \beta \triangleq \frac{\left|\bh^* \bv \right|}{ \| \bh\|}$.
A lower bound of the rate is
\begin{eqnarray} \label{eq:Capacity_MISO_fb_lb}
R_{\mr{MISO}}(1,B) &\geq & 2 \left(1 - \Hb \left(  Q\left( \sqrt{\gamma \|\bh \|^2 \cos^2 \beta \left(1 - \sin 2 |\theta| \right)}\right) \right) \right).
\end{eqnarray}
The channel capacity with perfect \ac{CSIT}, derived in \cite{Mo_Jianhua_TSP15}, is
\begin{equation} \label{eq:Capacity_MISO}
R_{\mr{MISO}}(1, \infty) = 2\left(1 - \Hb\left(  Q\left( \sqrt{\gamma \|\mb{h}\|^2}\right) \right)\right).
\end{equation}

Comparing \eqref{eq:Capacity_MISO_fb_lb} and \eqref{eq:Capacity_MISO}, maximizing the term ${\cos^2 \beta \left( 1 - \sin 2 |\theta| \right)}$ will minimize the power loss.
Averaging over the codebook $\mathcal{W}$ and the residual phase $\theta$, the power loss is
\begin{IEEEeqnarray}{rCl}
& & \mathbb{E}_{\mathcal{W}, \theta} \left[ \cos^2 \beta \left( 1 - \sin 2|\theta|\right) \right] \\
&\stackrel{(a)}{=}&\mathbb{E}_{\mathcal{W}} \left[ \cos^2 \beta \right] \mathbb{E}_{\theta} \left[ 1 - \sin 2|\theta| \right] \\
&\stackrel{(b)}{\geq}& \left( 1- 2^{-\frac{B_1}{\Nt-1}}\right)  \left( 1- {2^{-B_2}} \right) \label{eq:MISO_power_loss}
\end{IEEEeqnarray}
where $(a)$ follows by noting that $|\bh^*\bv|$ and $\angle \left( \bh^* \bv\right)$ are independent for RVQ,
$(b)$ follows from the facts $\mathbb{E}_{\mathcal{W}} \left[\cos^2 \beta \right] > 1 - 2^{ -\frac{B_1}{\Nt-1}}$ \cite[Lemma 1]{Jindal_IT06} and $\mathbb{E}_{\theta} \left[ 1 - \sin 2 |\theta| \right] > 1 - 2^{-B_2}$ proved in \eqref{eq:phase_quan_bound_loose}.
From \eqref{eq:MISO_power_loss}, it is seen that when $B_1=\Nt-1$ and $B_2=1$, the average power loss is at most $6$ $\mr{dB}$.

Another performance metric is rate loss, which may be more important than power loss. We now analyze the rate loss caused by limited feedback. Note that the rate of the quantized system saturates to $2$ $\mr{bps/Hz}$ at high SNR, which is a difference from the unquantized systems. The rate loss incurred by finite-rate feedback is
\begin{eqnarray}
\Delta R_{\mr{MISO}}(1,B) &=& R_{\mr{MISO}}(1, \infty) - R_{\mr{MISO}}(1,B)  \\
  &=& \Hb \left(  Q\left( \sqrt{\gamma \|\bh \|^2 \cos^2 \beta \left(1 - \sin 2 \theta \right)}\right) \right)  + \Hb \left(  Q\left( \sqrt{\gamma \|\bh \|^2 \cos^2 \beta \left(1 + \sin 2 \theta \right)}\right) \right) \nonumber \\
  & &  \quad - 2\Hb \left(  Q\left( \sqrt{\gamma \|\bh \|^2 }\right) \right) \\
&\leq& 2 - R_{\mr{MISO}}^{\mr{fb}} \\
&\leq& 2 \Hb \left(  Q\left( \sqrt{\gamma \|\bh \|^2 \cos^2 \beta \left(1 - \sin 2 |\theta| \right)}\right) \right).
\end{eqnarray}

To ensure $\Delta R_{\mr{MISO}}(1,B) \leq 2\epsilon$, it is required that
\begin{eqnarray}
\gamma \|\bh \|^2 \cos^2 \beta \left(1 - \sin 2 |\theta| \right) \geq \delta,
\end{eqnarray}
where $\Hb \left(  Q\left( \sqrt{\delta}\right) \right) = \epsilon$. Plugging in \eqref{eq:MISO_power_loss} and assuming $\mathbb{E}\left[\| \bh \|^2\right] = \Nt$, we obtain
\begin{eqnarray} \label{eq:MISO_high_SNR_fb}
\left( 1- 2^{-\frac{B_1}{\Nt-1}}\right)  \left( 1- {2^{-B_2}} \right) \geq \frac{\delta}{\gamma \Nt}.
\end{eqnarray}

In \eqref{eq:MISO_high_SNR_fb}, we see that given fixed rate loss, the required number of feedback bits actually {\it decreases} with the transmit signal power. This is in striking contrast with the unquantized MISO systems.

In Fig. \ref{fig:H_b_Q_sqrt_x}, it is shown that $\Hb \left( Q \left( \sqrt{5} \right) \right) \approx 0.1$. Therefore, if the numbers of feedback bits satisfy
\begin{eqnarray} \label{eq:MISO_high_SNR_fb_example}
\left( 1- 2^{-\frac{B_1}{\Nt-1}}\right)  \left( 1- {2^{-B_2}} \right) \geq \frac{5}{\gamma \Nt},
\end{eqnarray}
then the rate loss is less than $0.2$ $\mr{bps/Hz}$, or equivalently $90\%$ of the upper bound ($2$ $\mr{bps/Hz}$) is achieved.

%Assume perfect synchronization, the received signal is,
%\begin{align}
%	y = \bh^* \bv s + n,
%\end{align}
%
%The quantization output is then
%\begin{align}
%r = \mathcal{Q} \left(\bh^* \bv s + n\right),
%\end{align}
%where

\subsubsection{Multi-Bit Quantization}
Different from the case of 1-bit quantization where capacity-achieving QPSK signaling was adopted, we assume that Gaussian signaling is used at the transmitter. Although Gaussian signaling is suboptimal, it is amenable for analyses and close to optimal at low and medium SNR \cite{Mezghani_ISIT12, Mo_Jianhua_TSP15}.

By Bussgang's theorem \cite{Bussgang_52, Fletcher_JSTSP07, Mezghani_ISIT12}, the quantization output can be decoupled into two uncorrelated parts, i.e.,
\begin{IEEEeqnarray}{rCl}
r &=& (1-\eta_b) y + n_{\mr{Q}} \\
& =& (1-\eta_b) \bh^* \bv s + (1-\eta_b) n + n_{\mr{Q}},
\end{IEEEeqnarray}
where $\eta_b = \frac{\mathbb{E}[|r-y|^2]}{\mathbb{E}[|y|^2]}$ is the normalized mean squared error and $n_{\mathrm{Q}}$ is the quantization noise with variance $\sigma_{\mr{Q}}^2 = \eta_b (1-\eta_b) \mathbb{E}[|y|^2] = \eta_b(1-\eta_b) \left( | \bh^* \bv |^2 \Pt + \sigma_n^2 \right)$. Therefore, the effective noise $n_{\mathrm{ef}} \triangleq (1-\eta_b) n + n_{\mr{Q}}$ has variance $\eta_b(1-\eta_b) \left( | \bh^* \bv |^2 \Pt \right) + (1-\eta_b) \sigma_n^2 $. The values of $\eta_b \left(1\leq b \leq 8\right)$ are listed in Table \ref{tab:Eta_b}.
The resulting signal-to-quantization and noise ratio (SQNR) at the receiver is
\begin{IEEEeqnarray*}{rCl}
\mr{SQNR}
&=& \frac{ \left(1-\eta_b \right)^2 \Pt \left|\bh^*\bv \right|^2 }{ \left( 1-\eta_b \right)^2 \sigma_n^2 + \sigma_{\mr{Q}}^2 } = \frac{ \left(1-\eta_b \right) \gamma \left|\bh^*\bv \right|^2 }{ \eta_b  \gamma \left|\bh^*\bv \right|^2 + 1 }. \IEEEyesnumber
\end{IEEEeqnarray*}

Assuming that the noise $n_{\mathrm{Q}}$ follows the worst-case Gaussian distribution, the average achievable rate with perfect CSIT and conjugate beamforming is
%
%\begin{align}
%R =  \log_2 \left(1+ \frac{ \left(1-\eta_b \right) \Pt \left\|\bh\right\|^2 }{ \eta_b \Pt \left\|\bh \right\|^2 + \sigma_n^2} \right) < \log_2 \left( \frac{1}{\eta_b}\right)
%\end{align}
\begin{IEEEeqnarray}{rCl}
R_{\mr{MISO}}(b, \infty) &=& \mathbb{E}_{\bh} \left[\log_2 \left(1+ \frac{ \left(1-\eta_b \right) \gamma \left\|\bh\right\|^2 }{ \eta_b \gamma \left\|\bh \right\|^2 + 1} \right) \right] \\
&\stackrel{(a)}{\leq} & \log_2 \left(1+ \frac{ \left(1-\eta_b \right) \gamma \mathbb{E} \left[\left\|\bh\right\|^2 \right] }{ \eta_b \gamma \mathbb{E} \left[ \left\|\bh \right\|^2 \right] + 1} \right) \\
& \stackrel{(b)}{=} & \log_2 \left(1+ \frac{ \left(1-\eta_b \right) \gamma \Nt }{ \eta_b \gamma \Nt + 1} \right)
\end{IEEEeqnarray}
where $(a)$ follows from the concavity of the function $f(x)=\log_2 \left(1 + \frac{ax}{bx+c}\right) (a>0, b>0, c>0)$ when $x>0$, $(b)$ follows from the assumption of IID Rayleigh fading channel.

In the low and high SNR $\left(\frac{\gamma}{\sigma_n^2}\right)$ regimes, the average achievable rate with perfect CSIT is approximately,
\begin{align} \label{eq:SU_Rate_CSIT_low_high_SNR}
	R_{\mr{MISO}}(b, \infty) & \approx \left\{
	\begin{array}{lc}
	\log_2 \left(1+ \left(1-\eta_b \right) \gamma \Nt  \right), &\text{when $\gamma$ is small,} \\
	\log_2 \left( \frac{1}{\eta_b}\right), &\text{when $\gamma$ is large.}
	\end{array}
	\right.
\end{align}

It is seen that the high SNR rate is limited by the signal-to-quantization ratio (SQR) defined as $\mr{SQR} \triangleq \frac{1}{\eta_b}$.
Since $\eta_b \approx \frac{\pi \sqrt{3}}{2} 2^{-2b}$ when $b\geq 3$ \cite{Gersho_Book12}, the achievable rate at high SNR is
\begin{IEEEeqnarray}{rCl}
	R_{\mr{MISO}}(b, \infty)
	%&\approx & \log_2 \left( \frac{1}{\eta_b}\right) \\
	&\approx & 2 b - \log_2 \frac{\pi \sqrt{3}} {2} \\
	&\approx & 2  b - 1.44 \quad \text{bps/Hz}.
\end{IEEEeqnarray}
The values of $\log_2 \left( \frac{1}{\eta_b}\right)$ are also given in Table \ref{tab:Eta_b}.

Averaging over the RVQ codebooks, the achievable rate under limited feedback is
\begin{IEEEeqnarray}{rCl}
R_{\mr{MISO}}(b, B) 
&=& \mathbb{E}_{\bh, \mathcal{W}} \left[\log_2 \left(1+ \frac{ \left(1-\eta_b \right) \gamma \left|\bh^*\bv \right|^2 }{  \eta_b \gamma \left|\bh^*\bv \right|^2 + 1} \right) \right] \\
% &\stackrel{(a)}{\approx} \mathbb{E}_{\bh} \left[ \log_2 \left(1+ \frac{ \left(1-\eta_b \right) \Pt \left\|\bh \right\|^2 \left(1- 2^B \beta\left(2^B, \frac{\Nt}{\Nt-1}\right) \right) }{  \eta_b \Pt \left\|\bh \right\|^2 \left(1- 2^B \beta\left(2^B, \frac{\Nt}{\Nt-1}\right) \right)+\sigma_n^2} \right) \right] \\
&\stackrel{(a)}{\approx}  & \log_2 \left(1+ \frac{ \left(1-\eta_b \right) \gamma \Nt \left(1- 2^B \beta\left(2^B, \frac{\Nt}{\Nt-1}\right) \right) }{  \eta_b \gamma \Nt \left(1- 2^B \beta\left(2^B, \frac{\Nt}{\Nt-1}\right) \right)+1} \right) \IEEEeqnarraynumspace \\
&\stackrel{(b)}{\geq} & \log_2 \left(1+ \frac{ \left(1-\eta_b \right) \gamma \Nt \left(1- 2^{-\frac{B}{\Nt-1}} \right) }{  \eta_b \gamma \Nt \left(1- 2^{-\frac{B}{\Nt-1}} \right)+1} \right)
\end{IEEEeqnarray}
where $\beta(\cdot, \cdot)$ is a beta function. The approximation $(a)$ follows from $\mathbb{E}\left[\|\bh \|^2 \right] = \Nt$ and $\cos^2 \left(\angle \left(\bh, \bv\right) \right) = 1 - 2^B \beta\left(2^B, \frac{\Nt}{\Nt-1} \right)$ \cite{Au-Yeung_TWC07}, while
$(b)$ follows from the inequality $2^B \beta\left(2^B, \frac{\Nt}{\Nt-1}\right) \leq 2^{- \frac{B}{\Nt-1}}$ \cite{Jindal_IT06}.
%\begin{align}
%	\frac{\Nt}{\Nt-1} 2^{- \frac{B}{\Nt-1}} \leq 2^B \beta\left(2^B, \frac{\Nt}{\Nt-1}\right) \leq 2^{- \frac{B}{\Nt-1}}
%\end{align}

%To ensure that the rate loss is less than $\epsilon$, we have
%\begin{eqnarray}
%	B \geq f(\eta_b \gamma \epsilon)
%\end{eqnarray}

In the low and high SNR regimes, the average achievable rate with limited feedback is
\begin{align} \label{eq:SU_Rate_fb_low_high_SNR}
    R_{\mr{MISO}}(b, B) \approx\left\{
	\begin{array}{lc}
	\log_2 \left(1+  \left(1-\eta_b \right) \gamma \Nt \left( 1- 2^{-\frac{B}{\Nt-1}} \right)  \right), &\text{if $\gamma$ is small,} \\ %&\text{at low SNR,} \\
	\log_2 \left( \frac{1}{\eta_b}\right), & \text{if $\gamma$ is large.}  
	% &\text{at high SNR.}
	\end{array}
	\right.
\end{align}

%\begin{enumerate}
%\item That stick as a pointer was a mistake. You waved it around a lot, and it made an unpleasant scraping sound. In fact, you don't need to point all the much in general when you give a presentation. People will be able to follow most of the time. Only point when there is something really important.
%
%\item Be mindful of where you stand. You were in the projector light much of the time. You need to stand a little more still, away form the screen.
%
%\item You need to think much more carefully about what you are wearing during the presentation. You looked very sloppy. Wearing nice clothing is a sign of respect for the audience. You should at a minimum not wear running shoes, should have long pants, and a collared shirt. Even better is to have a sport jacket with it as well, and possibly a tie.
%\end{enumerate}

Comparing $R_{\mr{MISO}}(b, \infty)$ in \eqref{eq:SU_Rate_CSIT_low_high_SNR} and $R_{\mr{MISO}}(b, B)$ in \eqref{eq:SU_Rate_fb_low_high_SNR}, we find that at low SNR, the power loss between $R_{\mr{MISO}}(b, \infty)$ and $R_{\mr{MISO}}(b, B)$ is about $10 \log_{10} \left( 1- 2^{-\frac{B}{\Nt-1}} \right)$ dB. The result is similar to the case with infinite-bit ADCs \cite{Au-Yeung_TWC07, Mukkavilli_IT03}.
In contrast, at high SNR, both $R_{\mr{MISO}}(b, \infty)$ and $R_{\mr{MISO}}(b, B)$ approach the same upper bound and the rate loss due to limited feedback is zero.

The achievable rate with infinite-bit ADC and perfect CSIT is known as $R_{\mr{MISO}}(\infty, \infty)= \log_2 \left(1+ \Nt \gamma \right)$. We find that at low SNR, the power loss incurred by the finite-bit ADC is $10 \log_{10} \left(1 -\eta_b\right)$ dB while that by limited feedback is $10 \log_{10} \left(1 - 2^{-\frac{B}{\Nt-1}}\right)$ dB.

%\subsection{SIMO Channel with Limited Feedback}
%
%{\color{blue} We do not have the closed-form expression for the channel capacity.}
%
%Ideas:
%\begin{enumerate}
%	\item Only feed back the phase of each element in the channel vector $\bh \in \mathbb{C}^{\Nr \times 1}$. Based on these phases, design the transmit constellation, which is $4 \Nr$-PSK. See Fig. \ref{fig:SIMO_signaling} for the case $\Nr=3$ as an example.
%	
%	\item Use the lower bound provided by Bussgang theorem as an approximation at low SNR \cite{Mezghani_WSA07, Bai_Qing_ETT15}. {\color{blue}What does the receiver feedback?}
%	\begin{align} \label{eq:SIMO_AQNM}
%		R^{\mr{AQNM}} &= \log_2 \left( 1 + {\gamma} \mb{h}^* \mr{diag}\left\{ \frac{ 1 - \rho }{\sigma_n^2+ \rho \gamma |{h}_i|^2}\right\}_{i=1}^{\Nr} \mb{h}\right) \\
%		&= \log_2 \left( 1+  \sum_{i=1}^{\Nr} \frac{\gamma (1- \rho) |h_i|^2}{\sigma_n^2 + \gamma \rho |h_i|^2}\right)
%	\end{align}
%\end{enumerate}
%
%\begin{figure}[t]
%	\begin{centering}
%		\includegraphics[scale=.38]{Figures/SIMO_signaling.pdf}
%		\vspace{-0.1cm}
%		\centering
%		\caption{The transmitted symbols of a SIMO channel with 3 receive antennas. Here, $\angle h_1=0$ and $\angle h_2 \neq \angle h_3 \neq 0$.
%			The constellation contains 12 nonzero symbols and the symbol zero.}\label{fig:SIMO_signaling}
%	\end{centering}
%	\vspace{-0.5cm}
%\end{figure}

\subsection{MIMO Channel with Finite-Bit Quantization and Limited Feedback}
%\subsection{Zero-Forcing Precoding}
%Include the ZF results I did not included in the TSP paper
%\subsection{Single Stream Transmission}

In this section, we assume that single stream beamforming is used. For multi-stream transmission, zero-forcing precoding can be applied and the results are similar to the case of MU-MISO channel, which is presented in Section \ref{sec:MU_MISO}. Assuming that the beamforming vector is $\bv$, the received signal before quantization is
\begin{IEEEeqnarray}{rCl}
	\by &=& \bH \bv s + \bn.
\end{IEEEeqnarray}
By Bussgang's theorem \cite{Bussgang_52,Mezghani_ISIT12}, the quantized signal can be divided into two uncorrelated parts, 
\begin{IEEEeqnarray}{rCl}
	\br &=& \mathcal{Q} \left( \bH \bv s + \bn \right) \\
	&=& (1 - \eta_b) \bH \bv s + (1-\eta_b) \bn + \bn_{\mr{Q}}.	
\end{IEEEeqnarray}
where the covariance of $\bn_{\mathrm{Q}}$ is $\bC_{\bQ\bQ} = \bC_{\br\br}-\left(1-\eta_b\right)^2 \bC_{\by\by} = \bC_{\br\br}-\left(1-\eta_b\right)^2 \left(\bH \bv \bv^* \bH^* + \sigma_n^2 \bI \right)$. Therefore, assuming that $\bn_{\mathrm{Q}}$ follows the worst-case Gaussian distribution, the achievable rate is
\begin{IEEEeqnarray}{rCl}\label{eq:MIMO_Rate_lb}
	\overline{R}_{\mr{MIMO}}(b) = \log_{2} \left(\bI + \gamma (1-\eta_b)^2 \bv^* \bH^* \left(\bC_{\br\br}-\left(1-\eta_b\right)^2 \left(\bH \bv \bv^* \bH^* \right) \right)^{-1} \bH \bv \right).
\end{IEEEeqnarray}

\begin{figure}[t]
	\begin{centering}
		\includegraphics[width=0.8\columnwidth]{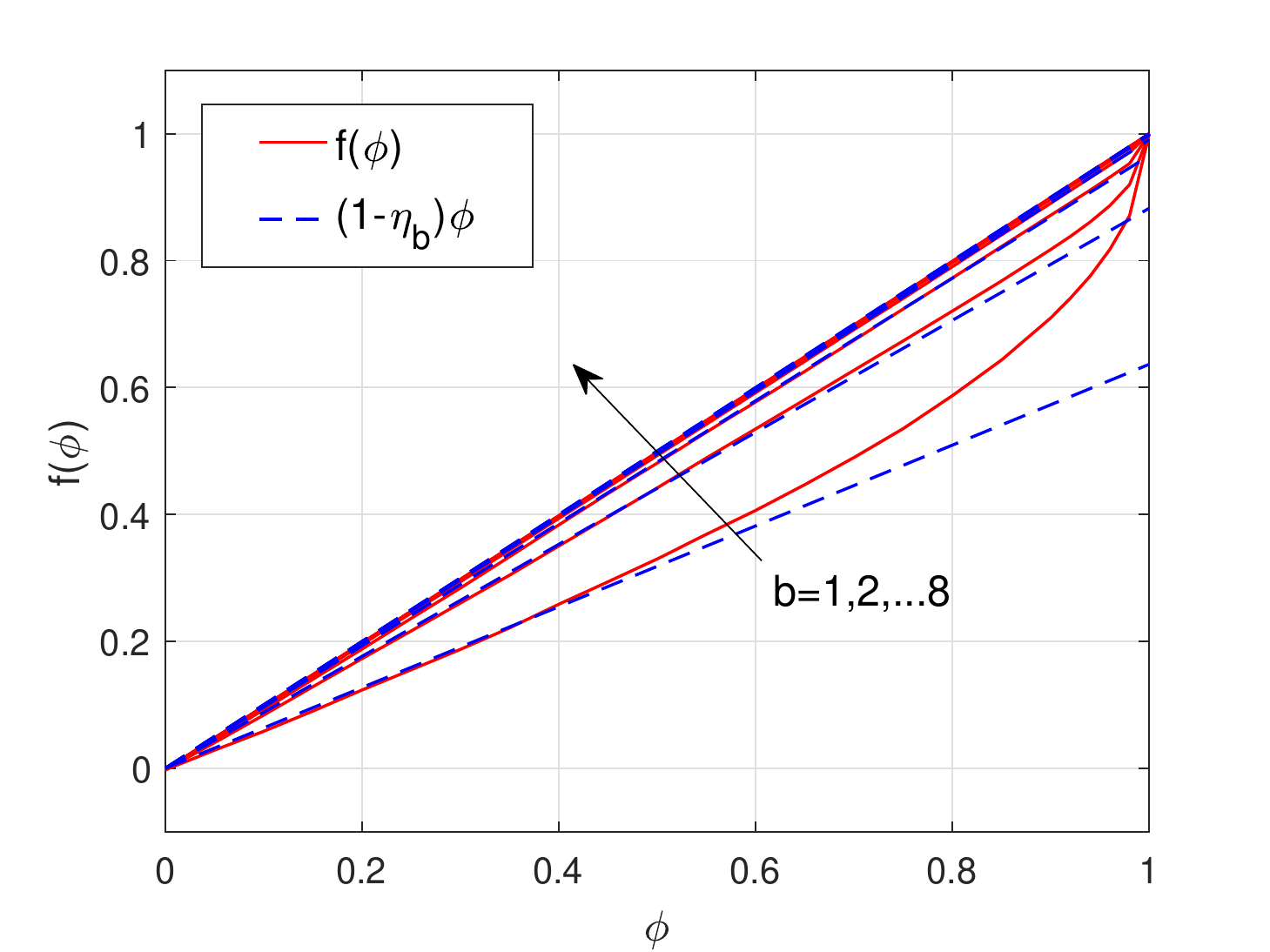}
		\vspace{-0.5cm}
		\centering
		\caption{This figure shows the impact of quantization on the correlation of signals. $\phi$ is the correlation coefficient of the input signal while $f(\phi)$ is the correlation coefficient of the $b$-bit ADC quantization output. The curves $(1-\eta_b) \phi$ for $b=1,2, \dots, 8$ are also shown for comparison. If $b=1$, it is seen the approximation is accurate when $\phi<0.5$. If $b\geq 2$, the approximation is quite accurate for the whole range of $\phi$. }\label{fig:f_function}
	\end{centering}
	\vspace{-0.3cm}
\end{figure}

Denote $\widehat{\by} \triangleq [\Re\left(\by\right)^T, \Im \left(\by \right)^T]^T$ and 
$\widehat{\br} \triangleq [\Re\left(\br\right)^T, \Im \left(\br \right)^T]^T$.
As $\widehat{\by}$ and $\widehat{\br}$ are both zero-mean, the covariance of $\hat{\br}$ can be written as
\begin{eqnarray}
\bC_{\widehat{\br}\widehat{\br}} &=& \diag{\bC_{\widehat{\br}\widehat{\br}}}^{\frac{1}{2}} \mathbf{\Phi}_{\widehat{\br}\widehat{\br}} \diag{\bC_{\widehat{\br}\widehat{\br}}}^{\frac{1}{2}} \\
&\stackrel{(a)}{=}& \left(1-\eta_b\right) \diag{\bC_{\widehat{\by}\widehat{\by}}}^{\frac{1}{2}} \mathbf{\Phi}_{\widehat{\br}\widehat{\br}} \diag{\bC_{\widehat{\by}\widehat{\by}}}^{\frac{1}{2}} \\
&=& \left(1-\eta_b\right) \diag{\bC_{\widehat{\by}\widehat{\by}}}^{\frac{1}{2}} f \left(\mathbf{\Phi}_{\widehat{\by}\widehat{\by}} \right) \diag{\bC_{\widehat{\by}\widehat{\by}}}^{\frac{1}{2}} \\
&=& \left(1-\eta_b\right) \diag{\bC_{\widehat{\by}\widehat{\by}}}^{\frac{1}{2}} f \left(\diag{\bC_{\widehat{\by}\widehat{\by}}}^{-\frac{1}{2}}\bC_{\widehat{\by}\widehat{\by}} \diag{\bC_{\widehat{\by}\widehat{\by}}}^{-\frac{1}{2}} \right) \diag{\bC_{\widehat{\by}\widehat{\by}}}^{\frac{1}{2}} \label{eq:R_rr}
\end{eqnarray}
where  $\mathbf{\Phi}_{\widehat{\by}\widehat{\by}}$ is the correlation matrix of $\widehat{\by}$, $(a)$ follows from the fact that $\diag{\bC_{\widehat{\br}\widehat{\br}}} = (1-\eta_b) \diag{\bC_{\widehat{\by}\widehat{\by}}}$. Here,
$f:[-1, 1] \rightarrow [-1, 1]$ denotes a function mapping from the correlation coefficient of two Gaussian input signals to the correlation coefficient of two output signals, i.e., 
\begin{IEEEeqnarray}{rCl}
	\phi_{\mathcal{Q}(x_m) \mathcal{Q}(x_n)} &=& f\left( \phi_{x_m x_n} \right),
\end{IEEEeqnarray}
where $x_m$ and $x_n$ are two Gaussian signals with correlation coefficient $\phi_{x_m x_n}$. Note that in \eqref{eq:R_rr}, $f(\cdot)$ is applied to each element of the matrix independently. $f(\cdot)$ can be computed by the following integration \cite{Price_IT58,Roth_DSP15}
\begin{eqnarray}
	f(\phi_{x_m x_n}) = \int_{x_m} \int_{x_n} \mathcal{Q}(x_m) \mathcal{Q}(x_n)  \Pr(x_m, x_n) \mr{d} x_n \mr{d} x_m
\end{eqnarray}
where $\Pr(x_m, x_n) = \frac{1}{2 \pi \sqrt{1-\phi_{x_m x_n}^2}} \exp \left(-\frac{1}{2(1-\phi_{x_m x_n}^2) } \left( x_m^2 + x_n^2 - 2 \phi_{x_m x_n} x_m x_n \right) \right)$ is the probability density function of two correlated Gaussian random variables. For a special case of 1-bit quantization, $f(\phi) = \frac{2}{\pi} \arcsin (\phi)$ \cite{Price_IT58}. \figref{fig:f_function} shows the values of $f(\cdot)$ for ADC resolution $1-8$ bits.
%A figure showing 
%$f(\cdot)$ for ADC resolution $1-4$ bits can be found in \cite[Figure 3]{Roth_DSP15}.

%
%draw a figure showing f for different bits. Fig. 2 in Roth_DSP15

%For a real-valued Gaussian vector $\widehat{\by}$ and its quantization output $\widehat{\br} = \sqrt{\frac{2}{\pi}} \mr{diag}\left(\widehat{\by} \widehat{\by}^T\right) \mr{sgn}\left({\widehat{\by}}\right)$, the correlation matrix can be written as
%\begin{eqnarray}
%\rho_{\widehat{\br}\widehat{\br}} = \frac{2}{\pi} \arcsin \left( \rho_{\widehat{\by}\widehat{\by}}\right)
%\end{eqnarray}

Denote $\by_r = \Re(\by)$, $\by_i = \Im(\by)$.
If $\by$ is circularly symmetric (which is the case in this paper since $s$ is a zero-mean complex Gaussian signal with independent real and imaginary parts), we have by circularly symmetry \cite[Section 2.6]{Proakis_Book08} that
\begin{IEEEeqnarray}{rCl}
	\bC_{\by_r \by_r} &=& \bC_{\by_i \by_i} = \frac{1}{2} \Re\left(\bC_{\by \by}\right), \\
	\bC_{\by_i \by_r} &=& - \bC_{\by_r \by_i} = \frac{1}{2} \Im \left(\bC_{\by \by}\right).
\end{IEEEeqnarray}
Therefore, the covariance of $\widehat{\by}$ is
\begin{eqnarray}
\bC_{\widehat{\by}\widehat{\by}} &=& \begin{bmatrix}
\bC_{\by_r \by_r} & \bC_{\by_r \by_i}\\
\bC_{\by_i \by_r} & \bC_{\by_i \by_i}
\end{bmatrix}  \\
&=& \begin{bmatrix}
\bC_{\by_r \by_r} & \bC_{\by_r \by_i}\\
-\bC_{\by_r \by_i} & \bC_{\by_r \by_r}
\end{bmatrix}
\end{eqnarray}
As a result, the covariance of the $\br_r=\Re(\br) \left(\br_i = \Im(\br) \right)$ is
\begin{IEEEeqnarray*}{rCl}
	& & \bC_{\br_r \br_r}  \\
	&=& \left(1-\eta_b\right) \diag{\bC_{\by_r \by_r}}^{\frac{1}{2}} f \left(\diag{\bC_{\by_r \by_r}}^{-\frac{1}{2}}\bC_{\by_r \by_r} \diag{\bC_{\by_r \by_r}}^{-\frac{1}{2}} \right) \diag{\bC_{\by_r \by_r}}^{\frac{1}{2}} \\
	&=& \frac{1-\eta_b}{2} \diag{\Re(\bC_{\by \by})}^{\frac{1}{2}} f \left(\diag{\Re(\bC_{\by \by})}^{-\frac{1}{2}}\Re(\bC_{\by \by}) \diag{\Re(\bC_{\by \by})}^{-\frac{1}{2}} \right) \diag{\Re(\bC_{\by \by})}^{\frac{1}{2}}\\
	&\stackrel{(a)}{=}& \frac{1-\eta_b}{2} \diag{\bC_{\by \by}}^{\frac{1}{2}} f \left(\diag{\bC_{\by \by}}^{-\frac{1}{2}} \Re(\bC_{\by \by}) \diag{\bC_{\by \by}}^{-\frac{1}{2}} \right) \diag{\bC_{\by \by}}^{\frac{1}{2}} \\
	&=& \bC_{\br_i \br_i}
\end{IEEEeqnarray*}
where $(a)$ follows from that all the diagonal elements of $\bC_{\by\by}$ are real numbers.

Similarly, the cross covariance of $\br_r$ and $\br_i$ is
\begin{IEEEeqnarray*}{rCl}
	& & \bC_{\br_i \br_r}\\
	&=& \left(1-\eta_b\right) \diag{\bC_{\by_r \by_r}}^{\frac{1}{2}} f \left(\diag{\bC_{\by_r \by_r}}^{-\frac{1}{2}}\bC_{\by_r \by_i} \diag{\bC_{\by_r \by_r}}^{-\frac{1}{2}} \right) \diag{\bC_{\by_r \by_r}}^{\frac{1}{2}} \\
	&=& \frac{1-\eta_b}{2} \diag{\Re(\bC_{\by \by})}^{\frac{1}{2}} f \left(\diag{\Re(\bC_{\by \by})}^{-\frac{1}{2}} \Im(\bC_{\by \by}) \diag{\Re(\bC_{\by \by})}^{-\frac{1}{2}} \right) \diag{\Re(\bC_{\by \by})}^{\frac{1}{2}} \\
	&=& \frac{1-\eta_b}{2} \diag{\bC_{\by \by}}^{\frac{1}{2}} f \left(\diag{\bC_{\by \by}}^{-\frac{1}{2}} \Im\left(\bC_{\by \by}\right) \diag{\bC_{\by \by}}^{-\frac{1}{2}} \right) \diag{\bC_{\by \by}}^{\frac{1}{2}} \\
	&=& -\bC_{\br_r \br_i}.
\end{IEEEeqnarray*}

As $\br$ is zero-mean and proper (i.e., $\bC_{\br_r \br_r} = \bC_{\br_i \br_i}$, $\bC_{\br_i \br_r} =-\bC_{\br_r \br_i}$), $\br$ is circularly symmetric \cite{Proakis_Book08}.
We conclude that for a Gaussian circularly symmetric signal, the quantized signal is still circularly symmetric, although the quantization is done separately on the real and imaginary parts of the input signal.

To sum up, the covariance of $\br$ is,
\begin{IEEEeqnarray}{rCl}
	\bC_{\br \br} &\stackrel{(a)}{=}& \bC_{\br_r \br_r} + \bC_{\br_i \br_i} + \j \left(\bC_{\br_i \br_r} - \bC_{\br_r \br_i}\right) \\
	&=& (1-\eta_b) \diag{\bC_{\by \by}}^{\frac{1}{2}} \left( f \left(\diag{\bC_{\by \by}}^{-\frac{1}{2}} \Re(\bC_{\by \by}) \diag{\bC_{\by \by}}^{-\frac{1}{2}} \right) \right. \nonumber \\
	& & \quad + \left. \j \, f \left(\diag{\bC_{\by \by}}^{-\frac{1}{2}} \Im\left(\bC_{\by \by}\right) \diag{\bC_{\by \by}}^{-\frac{1}{2}} \right) \right)  \diag{\bC_{\by \by}}^{\frac{1}{2}}
\end{IEEEeqnarray}
where $(a)$ can be verified from the definition of covariance matrix.

For the special case of 1-bit quantization, we have that
\begin{IEEEeqnarray}{rCl}
	\bC_{\br \br} &\stackrel{(a)}{=}& \bC_{\br_r \br_r} + \bC_{\br_i \br_i} + \j \left(\bC_{\br_i \br_r} - \bC_{\br_r \br_i}\right) \\
	&=& (1-\eta_b) \diag{\bC_{\by \by}}^{\frac{1}{2}} \left( \frac{2}{\pi} \arcsin \left(\diag{\bC_{\by \by}}^{-\frac{1}{2}} \Re(\bC_{\by \by}) \diag{\bC_{\by \by}}^{-\frac{1}{2}} \right) \right. \nonumber \\
	& & \quad + \left. \j \, \frac{2}{\pi} \arcsin \left(\diag{\bC_{\by \by}}^{-\frac{1}{2}} \Im\left(\bC_{\by \by}\right) \diag{\bC_{\by \by}}^{-\frac{1}{2}} \right) \right)  \diag{\bC_{\by \by}}^{\frac{1}{2}}.
\end{IEEEeqnarray}

For the general multi-bit case, $\bC_{\br\br}$ can only be computed by integration and a simple closed-form expression  is unavailable. 
For simplicity, we adopt the approximation that $f(\phi) \approx (1-\eta_b) \phi$ for $0\leq \phi < 1$. \figref{fig:f_function} compares the curves of $(1-\eta_b) \phi$ and $f(\phi)$. It is seen that the approximation is quite accurate when $\phi<0.5$ when $b=1$. Actually, the slope of $f(\phi_1) = \frac{2}{\pi} \arcsin(\phi)$ at $\phi=0$, which is $\frac{2}{\pi}$, is same as that of the approximation curve. As $b$ increases, the approximation becomes more accurate for large $\phi$.
By using the approximation $f(\phi) \approx (1-\eta_b) \phi$ for the non-diagonal elements and noticing that $f(1)=1$ for the diagonal elements, the covariance matrix of the quantization output is  
\begin{IEEEeqnarray}{rCl}
	\bC_{\br \br} 
%	&\approx & (1-\eta_b) \diag{\bC_{\by \by}}^{\frac{1}{2}} \left( (1-\eta_b) \left(\diag{\bC_{\by \by}}^{-\frac{1}{2}} \Re(\bC_{\by \by}) \diag{\bC_{\by \by}}^{-\frac{1}{2}} \right) \right. \nonumber \\
%	& & \quad + \left. \j \, (1-\eta_b) \left(\diag{\bC_{\by \by}}^{-\frac{1}{2}} \Im\left(\bC_{\by \by}\right) \diag{\bC_{\by \by}}^{-\frac{1}{2}} \right) \right)  \diag{\bC_{\by \by}}^{\frac{1}{2}} \\
	&\approx & (1-\eta_b) \left( \mr{diag}\left\{\bC_{\by \by} \right\} + (1-\eta_b) \mr{nondiag} \left\{\bC_{\by\by}\right\} \right).
\end{IEEEeqnarray}
which is same as that in \cite[Equation (28)]{Mezghani_ISIT12}.
Therefore, $\bC_{\bQ\bQ} \approx \eta_b (1-\eta_b) \mr{diag}\left\{\bC_{\by\by}\right\}= \eta_b \left(1-\eta_b\right) \left(\Pt \bH \bv \bv^* \bH^* + \sigma_n^2 \bI\right)$.
By assuming the quantization noise follows the worst-case Gaussian distribution, an achievable rate is
\begin{align} \label{eq:MIMO_Rate_approx_lb}
	\widetilde{R}_{\mr{MIMO}}(b, \infty) &\approx \log_2 \left( 1 + {\gamma} (1-\eta_b) \mb{v}^* \bH^* \left( \eta_b \gamma \mr{diag}\left\{ \bH \bv \bv^* \bH^* \right\}  + \bI \right)^{-1} \bH \bv \right).
\end{align}

Denote the $i$-th row of $\bH$ as $\bh_i^*$. The optimal choice of $\bv$ should be
\begin{eqnarray}
	\bv &=& \underset{\bv: \; \|\bv\|=1}{\arg \max} \; \mb{v}^* \bH^* \left( \eta_b \Pt \mr{diag}\left\{ \bH \bv \bv^* \bH^* \right\}  + \sigma_n^2 \bI \right)^{-1} \bH \bv\\
	&=& \underset{\bv: \; \|\bv\|=1}{\arg \max} \sum_{i=1}^{\Nr} \frac{(1-\eta_b) \gamma |\bh_i^* \bv|^2}{\eta_b \gamma |\bh_i^* \bv|^2+ 1} \\
	&=& \underset{\bv: \; \|\bv\|=1}{\arg \max} \sum_{i=1}^{\Nr} \frac{\bv^* \bA_i \bv}{\bv^* \bB_i \bv}
\end{eqnarray}
where $\bA_i = (1-\eta_b) \Pt \bh_i \bh_i^*$ and $\bB_i = \eta_b \Pt \bh_i \bh_i^* + \sigma_n^2 \bI_{\Nr}$. The optimization is in the form of maximizing a sum of generalized Rayleigh quotients. This problem is non-convex and finding the global optima is in general difficult \cite{Nguyen_16}. For the case with perfect CSIT, we assume that the transmitter chooses the best beamforming in the set $\left\{\mathcal{W}, \bv_{\mr{max}} \right\} $. If there is only finite-rate feedback, the receiver computes the term $\sum_{i=1}^{\Nr} \frac{(1-\eta_b) \gamma |\bh_i^* \bv|^2}{\eta_b \gamma |\bh_i^* \bv|^2+ 1}$ for each $\bv$ in the codebook $\mathcal{W}$ and feedbacks the index of the best one.

At low SNR, the quantization noise is dominated by AWGN, i.e., $\| \eta_b \Pt \mr{diag}\left\{ \bH \bv \bv^* \bH^* \right\} \|_F \ll \|\sigma_n^2 \bI \|_F$. As a result, the achievable rate is approximately to be
\begin{align} \label{eq:MIMO_AQNM_low_SNR}
	\widetilde{R}_{\mr{MIMO}}(b, \infty) &\approx \log_2 \left( 1 + \left(1-\eta_b \right) \gamma \mb{v}^* \bH^* \bH \bv \right).
\end{align}
Therefore, the optimal choice of the $\bv$ is $\bv_{\mr{max}}$. The receiver finds the vector of maximizing $\|\bH \bv\|$ and feeds back its index. Compared to the case without quantization, the SNR loss due to finite-resolution ADC is $10 \log_{10} (1-\eta_b)$ dB. In addition, the expected loss in SNR due to quantization resulting from codebook $\mathcal{W}$ can be written as
\begin{IEEEeqnarray}{rCl}
	\mathbb{E}_{\mathcal{W}}\left[ \sigma^2_{\mr{max}} - \underset{\bv \in \mathcal{W}}{\arg \max}\|\bH \bv \|^2 \right].
\end{IEEEeqnarray}
%
%At medium and high SNR, we propose to use an iterative algorithm to find $\bv$. 
%\begin{align}
%	\bv^{(n+1)} = \bv_{\mr{max}} \left( \left( \eta_b \gamma \mr{diag} \left\{\bH \bv \bv^* \bH^* \right\} + \sigma_n^2 \bI \right)^{-\frac{1}{2}} \bH \right)
%\end{align}

At high SNR, the achievable rate converges to
\begin{IEEEeqnarray}{rCl}
	\widetilde{R}_{\mr{MIMO}}(b, \infty) &\approx & \log_2 \left(1+ \sum_{i=1}^{\Nr} \frac{(1-\eta_b) \gamma |\bh_i^* \bv|^2}{\eta_b \gamma |\bh_i^* \bv|^2+ 1}\right) \\
	& \approx & \log_2 \left(1+ \frac{1-\eta_b}{\eta_b}\Nr \right) \quad \text{when  $\gamma$ is large,}
\end{IEEEeqnarray}
regardless of the choice of beamforming vector $\bv$. Therefore, there is no rate loss between the two cases of perfect CSIT and finite-bit feedback. We also find that the rate increases logarithmically with the number of receiver antennas.

\section{Multi-User MISO Channel with Finite-bit ADCs and Limited Feedback}\label{sec:MU_MISO}

\begin{figure}[t]
	\begin{centering}
		\includegraphics[width = 0.8\columnwidth]{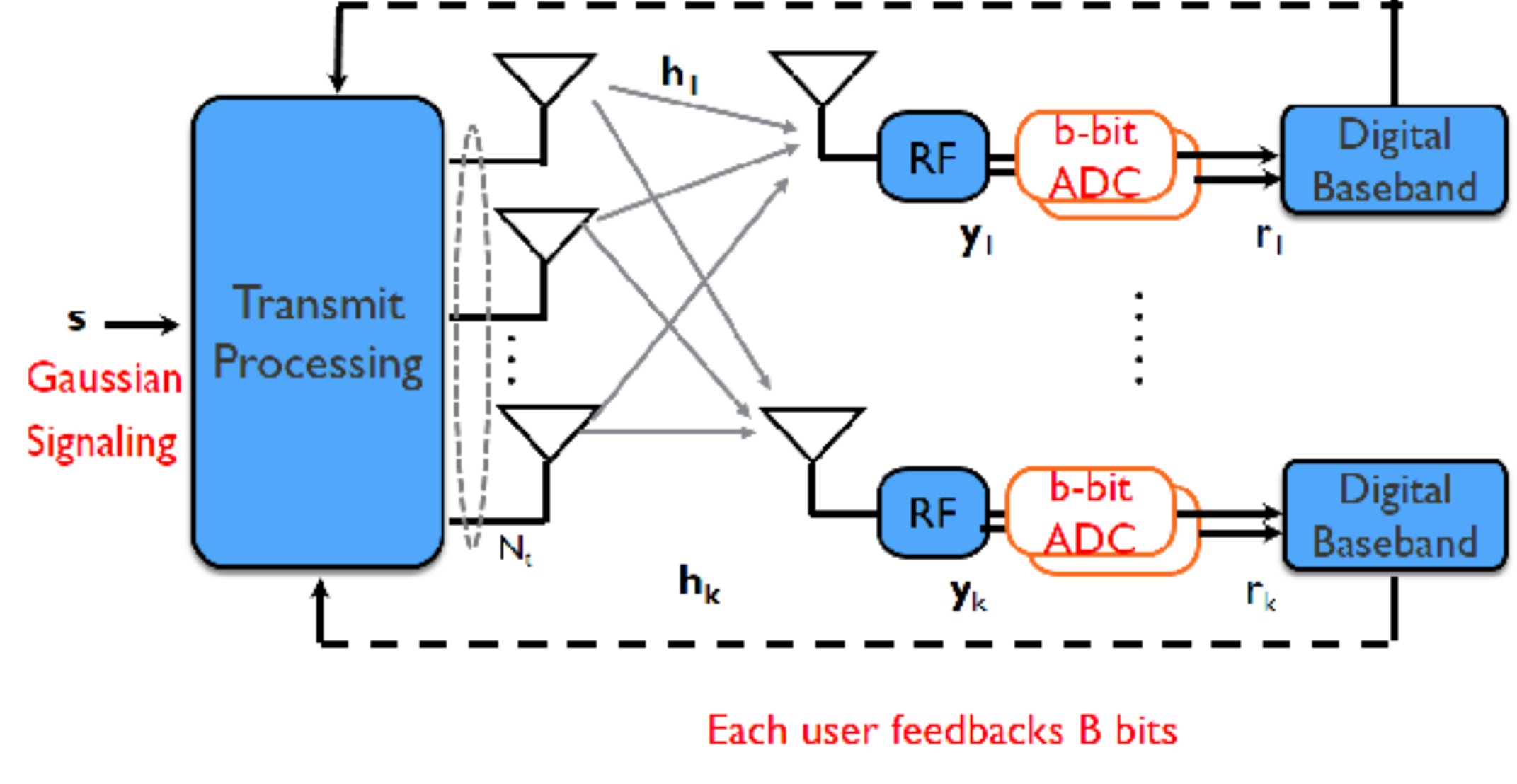}
		\vspace{-0.5cm}
		\centering
		\caption{A multi-user MISO system with finite-bit quantization and limited feedback. At each receiver, there are two $b$-bit ADCs. There is also a low-rate feedback path from each receiver to the transmitter. Note that there is no limitation on the structure of the transmitter.}\label{fig:MU_MISO}
	\end{centering}
	\vspace{-0.3cm}
\end{figure}

%\subsection{Single-Bit Quantization}
%
%\subsection{Multi-Bit Quantization}

We now consider a multi-user MISO channel shown in \figref{fig:MU_MISO} where a $\Nt$-antenna transmitter sends signals to $K \left(1<K \leq \Nt\right)$ single-antenna receivers.
The quantization output at the $k$-th receiver is
\begin{IEEEeqnarray}{rCl}
	r_k & =& \mathcal{Q} \left( \sqrt{\frac{\Pt}{K}} \sum_{i=1}^{K}\bh_k^*\bv_i s_i  + n_k \right) \\
	%& = (1-\eta_{b}) \left( \sqrt{\rho} \sum_{i=1}^{i=K}\bh_k^*\bv_i s_i  + n_k \right) + n_{\mr{Q}} \\
	& =&  (1-\eta_{b}) \sqrt{\frac{\Pt}{K}} \bh_k^*\bv_i s_i + (1-\eta_{b}) \sqrt{\frac{\Pt}{K}} \sum_{i=1, i\neq k}^{K}\bh_k^*\bv_i s_i + (1-\eta_{b}) n_k +n_{\mr{Q}},
\end{IEEEeqnarray}
where $\frac{\Pt}{K}$ is the power allocated to each user, $\bv_i$ is the beamforming vector for user $i$, $n_k \sim \mathcal{CN}(0, \sigma_n^2)$ is the circularly symmetric complex Gaussian noise, and the quantization noise $n_{\mr{Q}}$ has variance $\eta_{b} (1-\eta_{b})\left( \frac{\Pt}{K} \sum_{i=1, i\neq k}^K \left|\bh_k^*\bv_i \right|^2 +\sigma_n^2\right)$. Therefore, the signal-to-interference, quantization and noise ratio (SIQNR) at the $k$-th receiver is
\begin{IEEEeqnarray*}{rCl}
	\mr{SIQNR}_k
	%&=& \frac{ \left(1-\eta_b \right)^2 \rho \left|\bh_k^*\bv_i \right|^2 }{ \left( 1-\eta_b \right) \left( \rho \sum_{i=1, i \neq k}^K \left|\bh_k^*\bv_i \right|^2 +\sigma_n^2\right) + \eta_b (1-\eta_b) \rho \left|\bh_k^*\bv_k \right|^2}\\
	&=& \frac{ \left(1-\eta_b \right) \rho \left|\bh_k^*\bv_i \right|^2 }{  \eta_b \rho \left|\bh_k^*\bv_k \right|^2 + \rho \sum_{i=1, i\neq k}^K \left|\bh_k^*\bv_i \right|^2 +1}, \IEEEyesnumber
\end{IEEEeqnarray*}
where $\rho \triangleq \frac{\Pt}{ K \sigma_n^2} = \frac{\gamma}{K}$.

%\begin{align}
%  R_k = \log_2 \left(1+ \mr{SINR}_k\right)
%\end{align}

%\begin{align}
%  R_k^{\mr{ZF}}
%  &= \log_2 \left(1+ \frac{ \left(1-\eta_b \right) \rho \left|\bh_k^*\bv_k^{\mr{ZF}} \right|^2 }{ \eta_b \rho \left|\bh_k^*\bv_k^{\mr{ZF}} \right|^2 + \sigma_n^2} \right)
%\end{align}
%

If there is perfect CSIT and the transmitter designs zero-forcing beamforming $\bv_i^{\mr{ZF}} \left(1\leq i \leq K\right)$ such that $\bh_k^* \bv_i^{\mr{ZF}}=0$ for $k\neq i$, the average rate per user is
\begin{IEEEeqnarray}{rCl}
	R_{\mr{MISO}}^{\mr{ZF}}(b, \infty)
	&=& \mathbb{E}_{\bH} \left[\log_2 \left(1+ \frac{ \left(1-\eta_b \right) \rho \left|\bh_k^*\bv_k^{\mr{ZF}} \right|^2 }{ \eta_b \rho \left|\bh_k^*\bv_k^{\mr{ZF}} \right|^2 + 1} \right) \right] \\
	&\stackrel{(a)}{\leq}& \log_2 \left(1+ \frac{ \left(1-\eta_b \right) \rho (\Nt-K+1) }{ \eta_b \rho (\Nt-K+1) + 1} \right)
	%&=& \log_2 \left(\frac{ \rho (\Nt-K+1) + \sigma_n^2 }{ \eta_b \rho (\Nt-K+1) + \sigma_n^2} \right)
\end{IEEEeqnarray}
where $\mathbb{E}_{\bH}\left[\left|\bh_k^* \bv_k^{\mr{ZF}}\right|^2\right] = \mathbb{E}_{\bH}\left[\|\bh_k\|^2\right] \mathbb{E}_{\bH}\left[ |\widetilde{\bh}_k^*  \bv_k^{\mr{ZF}} |^2 \right] = \mathbb{E}_{\bH} \left[\|\bh_k\|^2\right] \mathbb{E}_{\bH} \left[ \cos ^2 \angle \left(\widetilde{\bh}_k, \bv_k^{\mathrm{ZF}}\right)\right]
=\Nt \frac{\Nt-K+1}{\Nt} = \Nt-K+1$ where $\widetilde{\bh}_k = \frac{\bh}{\|\bh\|}$.

We assume that $B$  bits are used to convey the channel direction information. A codebook $\mathcal{W} = \left\{\widehat{\bh}^{(0)}, \widehat{\bh}^{(1)}, \cdots, \widehat{\bh}^{(2^{B}-1)} \right\}$ is shared by the transmitter and receiver. The receiver sends back the index of the optimal codeword from the codebook. Then the transmitter performs beamforming based on the feedback information. Similar to a system with infinite-resolution ADCs, random vector quantization (RVQ) is also adopted to quantize the direction of channel $\bh$. %In the codebook $\mathcal{W}$, each of the quantization vectors is independently chosen from the isotropic distribution on the Grassmannian manifold $\mathcal{G}(\Nt, 1)$ \cite{Love_IT03}.

In the case without perfect CSIT, each receiver feeds back $B$ bits as the index of the quantized channel $\widehat{\bh}_k$, then the transmitter designs zero-forcing precoding based on $\widehat{\bh}_k \left(1\leq k \leq K\right)$. The average achievable rate is
\begin{IEEEeqnarray}{rCl}
		& & R_{\mr{MISO}}^{\mr{ZF}}\left(b, B\right)\\
		& = & \mathbb{E}_{\bH, \mathcal{W}} \left[\log_2 \left(1+ \frac{ \left(1-\eta_b \right) \rho \left|\bh_k^*\bv_k \right|^2 }{  \eta_b \rho \left|\bh_k^*\bv_k \right|^2 + \rho \sum_{i=1, i\neq k}^K \left|\bh_k^*\bv_i \right|^2 +1} \right) \right] \label{eq:MU_Rate_fb} \\
		& \approx & \log_2 \left(1+ \frac{ \left(1-\eta_b \right) \rho \mathbb{E} \left[ \left|\bh_k^*\bv_k \right|^2 \right] }{  \eta_b \rho \mathbb{E} \left[ \left|\bh_k^*\bv_k \right|^2 \right] + \rho \sum_{i=1, i\neq k}^K \mathbb{E} \left[ \left|\bh_k^*\bv_i \right|^2 \right] +1} \right) \\
		& = & \log_2 \left(1+ \frac{ \left(1-\eta_b \right) \rho \left(\Nt-K+1\right) \left( 1-2^B \beta\left(2^B, \frac{\Nt}{\Nt-1}\right) \right) }{  \eta_b \rho \left(\Nt-K+1\right) \left( 1-2^B \beta\left(2^B, \frac{\Nt}{\Nt-1}\right) \right) + \rho \left(K-1\right)  \frac{\Nt}{\Nt-1} 2^B \beta\left(2^B, \frac{\Nt}{\Nt-1}\right) +1} \right) \nonumber \\
		& \geq & \log_2 \left(1+ \frac{ \left(1-\eta_b \right) \rho \left(\Nt-K+1\right) \left( 1-2^{-\frac{B}{\Nt-1}} \right) }{  \eta_b \rho \left(\Nt-K+1\right) \left( 1-2^{-\frac{B}{\Nt-1}} \right) + \rho \left(K-1\right)  \frac{\Nt}{\Nt-1} 2^{-\frac{B}{\Nt-1}} +1} \right) \label{eq:MU_Rate_fb_LB}
		%& = & \log_2 \left(\frac{ \rho \left(\Nt-K+1\right) \left( 1-2^B \beta\left(2^B, \frac{\Nt}{\Nt-1}\right) \right) + \rho \left(K-1\right)  \frac{\Nt}{\Nt-1} 2^B \beta\left(2^B, \frac{\Nt}{\Nt-1}\right) +\sigma_n^2} {  \eta_b \rho \left(\Nt-K+1\right) \left( 1-2^B \beta\left(2^B, \frac{\Nt}{\Nt-1}\right) \right) + \rho \left(K-1\right)  \frac{\Nt}{\Nt-1} 2^B \beta\left(2^B, \frac{\Nt}{\Nt-1}\right) +\sigma_n^2} \right)
\end{IEEEeqnarray}

%\begin{align}
%	&R^{\mr{ZF,fb}} \\
%	={} & \mathbb{E}_{\bH, \mathcal{W}} \left[\log_2 \left(1+ \frac{ \left(1-\eta_b \right) \rho \left|\bh_k^*\bv_k \right|^2 }{  \eta_b \rho \left|\bh_k^*\bv_k \right|^2 + \rho \sum_{i=1, i\neq k}^K \left|\bh_k^*\bv_i \right|^2 +\sigma_n^2} \right) \right] \\
%	\geq {} & \log_2 \left(1+ \frac{ \left(1-\eta_b \right) \rho \left(\Nt-K+1\right) \left( 1-2^B \beta\left(2^B, \frac{\Nt}{\Nt-1}\right) \right) }{  \eta_b \rho \left(\Nt-K+1\right) \left( 1-2^B \beta\left(2^B, \frac{\Nt}{\Nt-1}\right) \right) + \rho \left(K-1\right)  \frac{\Nt}{\Nt-1} 2^B \beta\left(2^B, \frac{\Nt}{\Nt-1}\right) +\sigma_n^2} \right) \\
%	= {} & \log_2 \left(\frac{ \rho \left(\Nt-K+1\right) \left( 1-2^B \beta\left(2^B, \frac{\Nt}{\Nt-1}\right) \right) + \rho \left(K-1\right)  \frac{\Nt}{\Nt-1} 2^B \beta\left(2^B, \frac{\Nt}{\Nt-1}\right) +\sigma_n^2} {  \eta_b \rho \left(\Nt-K+1\right) \left( 1-2^B \beta\left(2^B, \frac{\Nt}{\Nt-1}\right) \right) + \rho \left(K-1\right)  \frac{\Nt}{\Nt-1} 2^B \beta\left(2^B, \frac{\Nt}{\Nt-1}\right) +\sigma_n^2} \right)
%\end{align}

In \eqref{eq:MU_Rate_fb}-\eqref{eq:MU_Rate_fb_LB}, we use the equality $\mathbb{E}_{\bH, \mathcal{W}} \left[ \left|\bh_k^*\bv_i \right|^2 \right] = \mathbb{E}_{\bH} \left[ \left\|\bh_k \right\|^2 \right] \mathbb{E}_{\bH, \mathcal{W}} \left[ \left|\widetilde{\bh}_k^*\bv_i \right|^2 \right] = \frac{\Nt}{\Nt-1} 2^B \beta\left(2^B, \frac{\Nt}{\Nt-1}\right)$ \cite{Jindal_IT06} and the lower bound of $\mathbb{E}_{\bH, \mathcal{W}} \left[ \left| \bh_k^* \bv_k \right|^2 \right]$ as follows.
\begin{IEEEeqnarray*}{rCl}
	\mathbb{E} \left[ \left| \bh_k^* \bv_k \right|^2 \right]
	&\geq & \mathbb{E} \left[ \left| \bh_k^* \widehat{\bh}_k \right|^2 \right] \mathbb{E} \left[ \left| \widehat{\bh}_k^* \bv_k \right|^2 \right] \\
	&=& \mathbb{E} \left[ \left| \bh_k \right|^2 \right] \mathbb{E} \left[ \left| \widetilde{\bh}_k^* \widehat{\bh}_k \right|^2 \right] \mathbb{E} \left[ \left| \widehat{\bh}_k^* \bv_k \right|^2 \right] \IEEEyesnumber \\
	&\stackrel{(a)}{=}& \Nt \left( 1-2^B \beta\left(2^B, \frac{\Nt}{\Nt-1}\right) \right) \frac{\Nt-K+1}{\Nt},
\end{IEEEeqnarray*}
where $(a)$ follows from the equalities $\mathbb{E} \left[ \left| \widetilde{\bh}_k^* \widehat{\bh}_k \right|^2 \right] = 1-2^B \beta\left(2^B, \frac{\Nt}{\Nt-1}\right)$ \cite{Au-Yeung_TWC07} and $\mathbb{E} \left[ \left| \widehat{\bh}_k^* \bv_k \right|^2 \right] = \frac{\Nt - K+1}{\Nt}$.

Therefore, the rate loss incurred by limited feedback is
\begin{IEEEeqnarray}{rCl}
	\Delta R_{\mr{MISO}}^{\mathrm{ZF}}(b) & = & R_{\mr{MISO}}^{\mr{ZF}}\left(b, \infty \right) - R_{\mr{MISO}}^{\mr{ZF}}\left(b, B \right),
\end{IEEEeqnarray}
and has an upper bound
	\begin{IEEEeqnarray*}{rCl} \label{eq:MU_Rate_loss_ub}
		\Delta \overline{R}_{\mr{MISO}}^{\mathrm{ZF}}(b) & = & \log_2 \left(1+ \frac{ \left(1-\eta_b \right) \rho (\Nt-K+1) }{ \eta_b \rho (\Nt-K+1) + 1} \right) \\
		& & - \log_2 \left(1+ \frac{ \left(1-\eta_b \right) \rho \left(\Nt-K+1\right) \left( 1-2^{-\frac{B}{\Nt-1}} \right) }{  \eta_b \rho \left(\Nt-K+1\right) \left( 1-2^{-\frac{B}{\Nt-1}} \right) + \rho \frac{\left(K-1\right) \Nt}{\Nt-1} 2^{-\frac{B}{\Nt-1}} +1} \right). \IEEEyesnumber
	\end{IEEEeqnarray*}

When the SNR $\left(\frac{\Pt}{\sigma_n^2} =
K \rho \right)$ is low, the performance loss is
\begin{IEEEeqnarray*}{rCl} \label{MU_MISO_Rate_loss_low_SNR}
	\Delta \overline{R}_{\mr{MISO}}^{\mathrm{ZF}} (b) 
	& \approx &  \log_2 \left(1+  \left(1-\eta_b \right) \rho (\Nt-K+1)  \right)  \\
	& & - \log_2 \left(1+  \left(1-\eta_b \right) \rho \left(\Nt-K+1\right) \left( 1-2^{-\frac{B}{\Nt-1}} \right)  \right). \IEEEyesnumber
\end{IEEEeqnarray*}
It is found there is a power loss $\approx 10 \log_{10} \left( 1-2^{-\frac{B}{\Nt-1}} \right)$ dB which is similar to the single-user case shown in Section \ref{sec:SU_MISO}.

At high SNR, the rate loss is
\begin{IEEEeqnarray}{rCl}
	\Delta \overline{R}_{\mr{MISO}}^{\mr{ZF}}(b) & \approx & \log_2 \left( 1 + \frac{1 - \eta_b}{\eta_b } \frac{1}{\frac{C_1}{C_2} + 1}\right)
\end{IEEEeqnarray}
where $C_1 \triangleq \left(\Nt-K+1\right) \left( 1-2^{-\frac{B}{\Nt-1}} \right)$  and $C_2 \triangleq \frac{\left(K-1\right) \Nt}{\Nt-1} 2^{-\frac{B}{\Nt-1}}$. To guarantee that the rate loss is less than $D$, the number of feedback bits $B$ should be large enough such that $\frac{1 - \eta_b}{\eta_b } \frac{1}{C_1/C_2+1}<2^D-1$.

When $b \geq 3$, $1-\eta_b \approx 0$ as shown in Table \ref{tab:Eta_b}. If $B \gg \Nt-1$,
\begin{align} 
	\frac{C_1}{C_2} +1 \approx \frac{(\Nt-K+1)(\Nt-1)}{\Nt (K-1)} 2^{\frac{B}{\Nt-1}}.
\end{align}
In this case, to keep the rate loss constant, we want the following term
\begin{align} \label{eq:MU_MISO_scaling_law}
	\frac{1}{\eta_b 2^{\frac{B}{\Nt-1}}}  \approx  \frac{2}{\pi \sqrt{3}} 2^{2b} 2^{-\frac{B}{\Nt-1}}
	=  \frac{2}{\pi \sqrt{3}} 2^{2 \left( b - \frac{B}{2 \left( \Nt-1 \right)} \right)}
\end{align}
to be less than a constant.
Therefore, if the ADC resolution $b$ increase $1$ bit, the number of feedback bits $B$ should increase by $2(\Nt-1)$.

\begin{figure}[t]
	\begin{centering}
		\includegraphics[width=0.7\columnwidth]{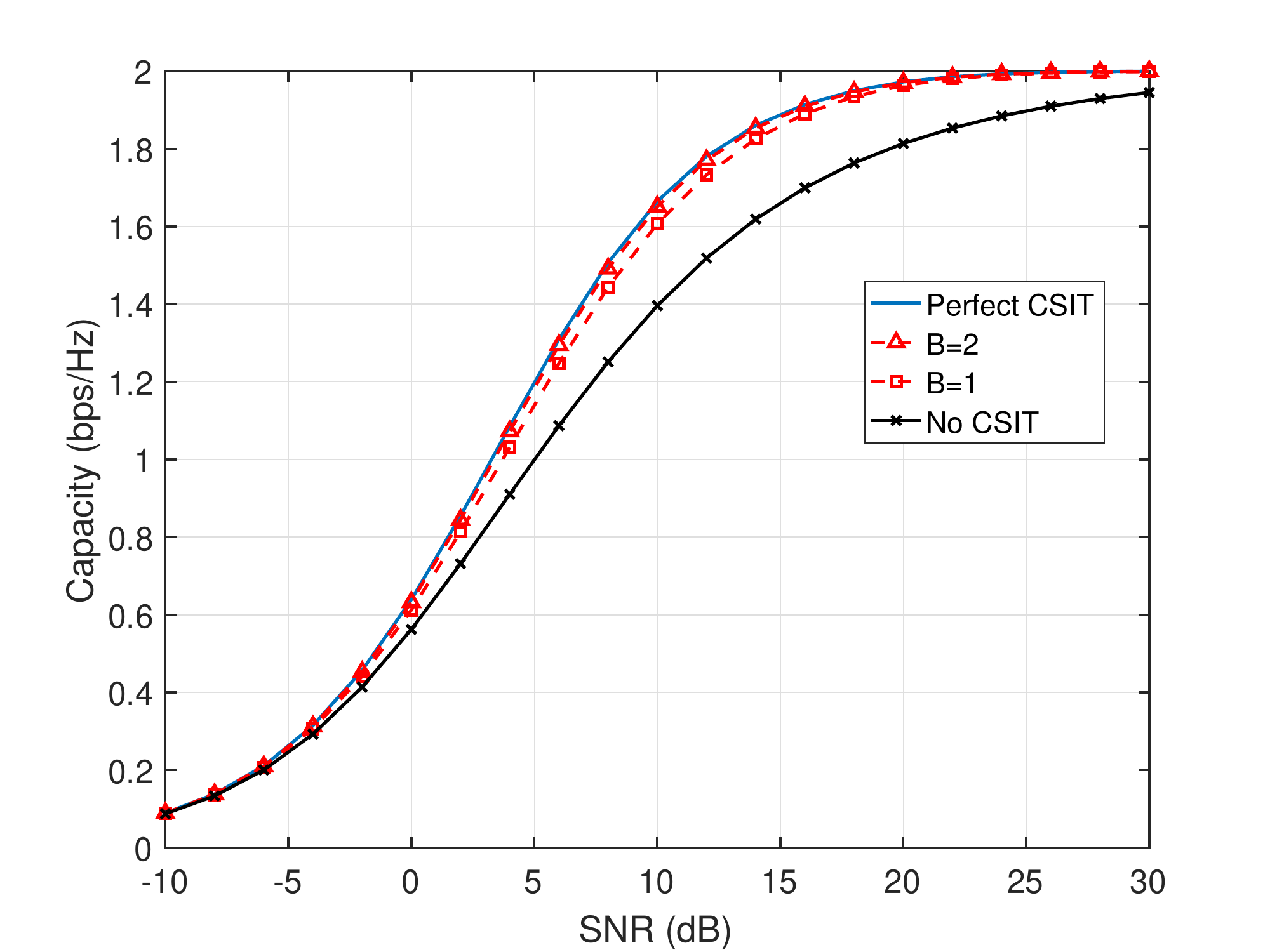}
		\vspace{-0.1cm}
		\centering
		\caption{The rate of a SISO system with CSIT, no CSIT and limited feedback 
			when the ADC resolution is 1-bit.
			%The channel $h$ experiences Rayleigh fading, i.e., $h \sim \mathcal{CN}(0, 1).$
		}\label{fig:SISO_Capacity}
	\end{centering}
	\vspace{-0.3cm}
\end{figure}

\begin{figure}[t]
	\centering
	\subfigure[$\Nt=4$]{
		\includegraphics[width=0.7\columnwidth]{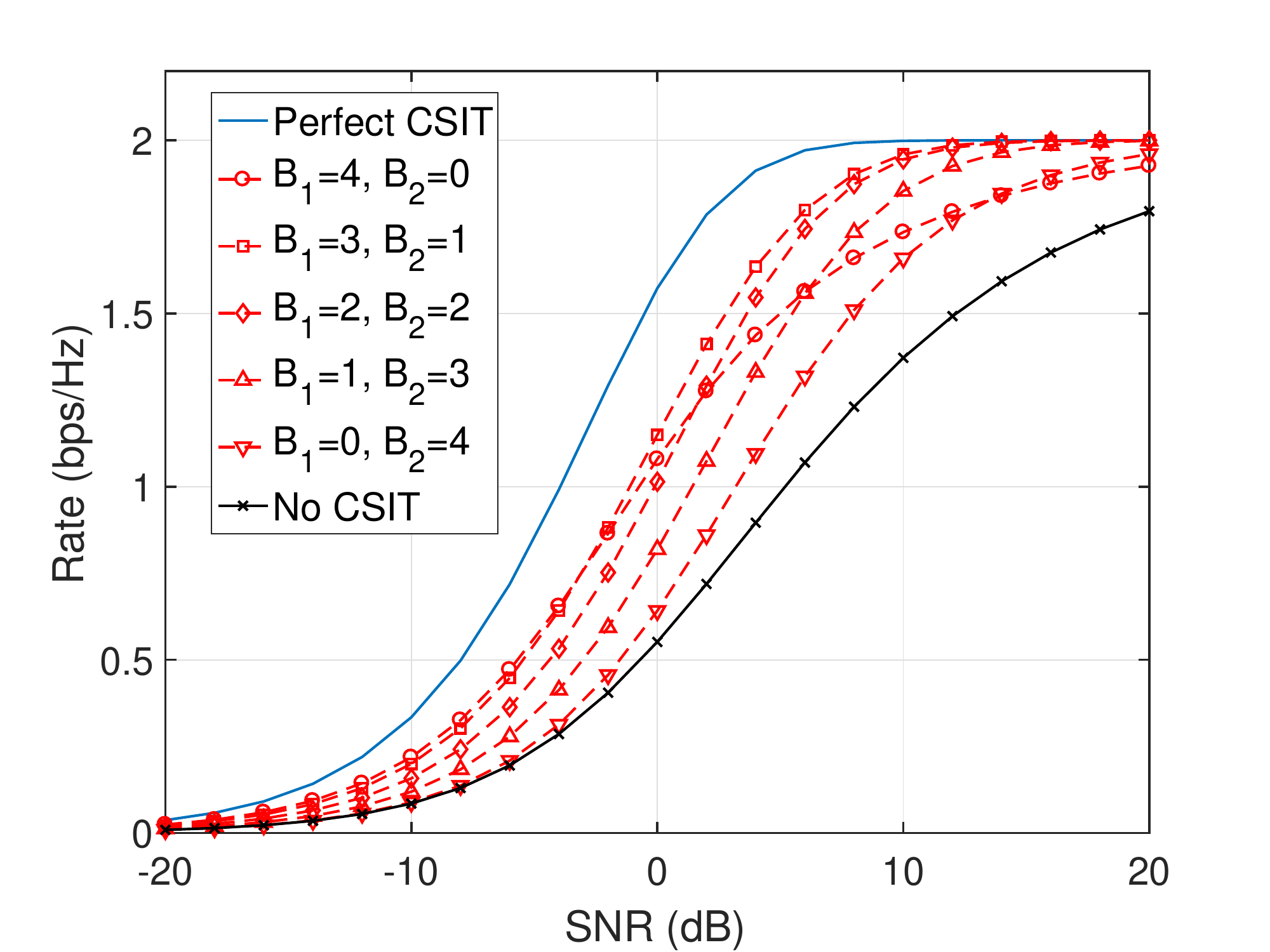}
		\label{fig:MISO_Capacity_Nt_4}
	}
	\subfigure[$\Nt=16$]{
		\includegraphics[width=0.7\columnwidth]{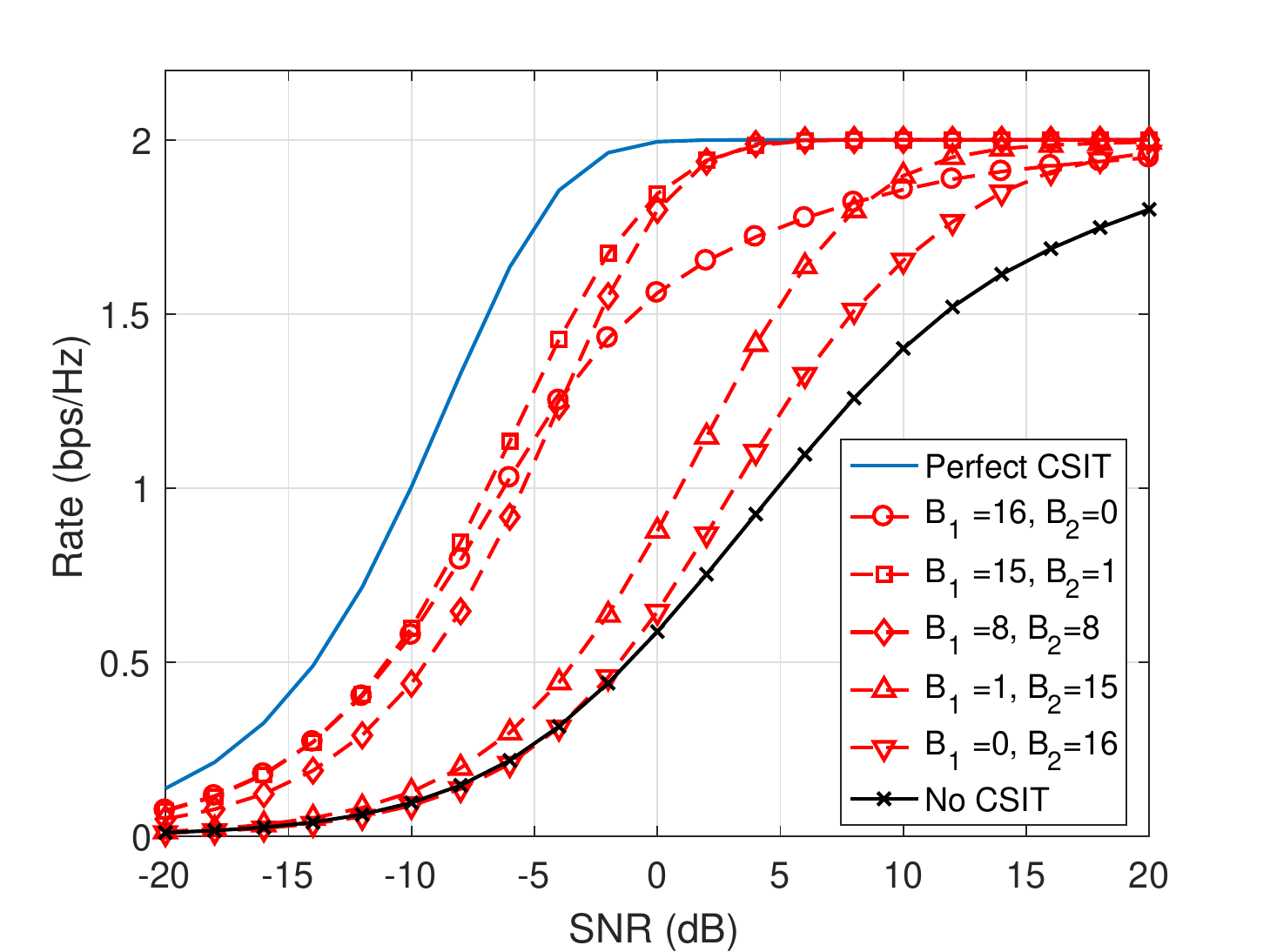}
		\label{fig:MISO_Capacity_Nt_16}
	}
	\caption{The achievable rate of a MISO system with CSIT, no CSIT and limited feedback when the ADC resolution is 1-bit. Two different cases, $\Nt=4$ and $\Nt=16$ are shown. The total number of feed back bits, i.e., $B=B_1+B_2$, is equal to $\Nt$. The best way to assigning the feed back bits is `$B_1 = \Nt-1, B_2=1$'.}
	\label{fig:MISO_Capacity_1bit_ADC}
\end{figure}

\begin{figure}[t]
	\begin{centering}
		\includegraphics[width=0.7\columnwidth]{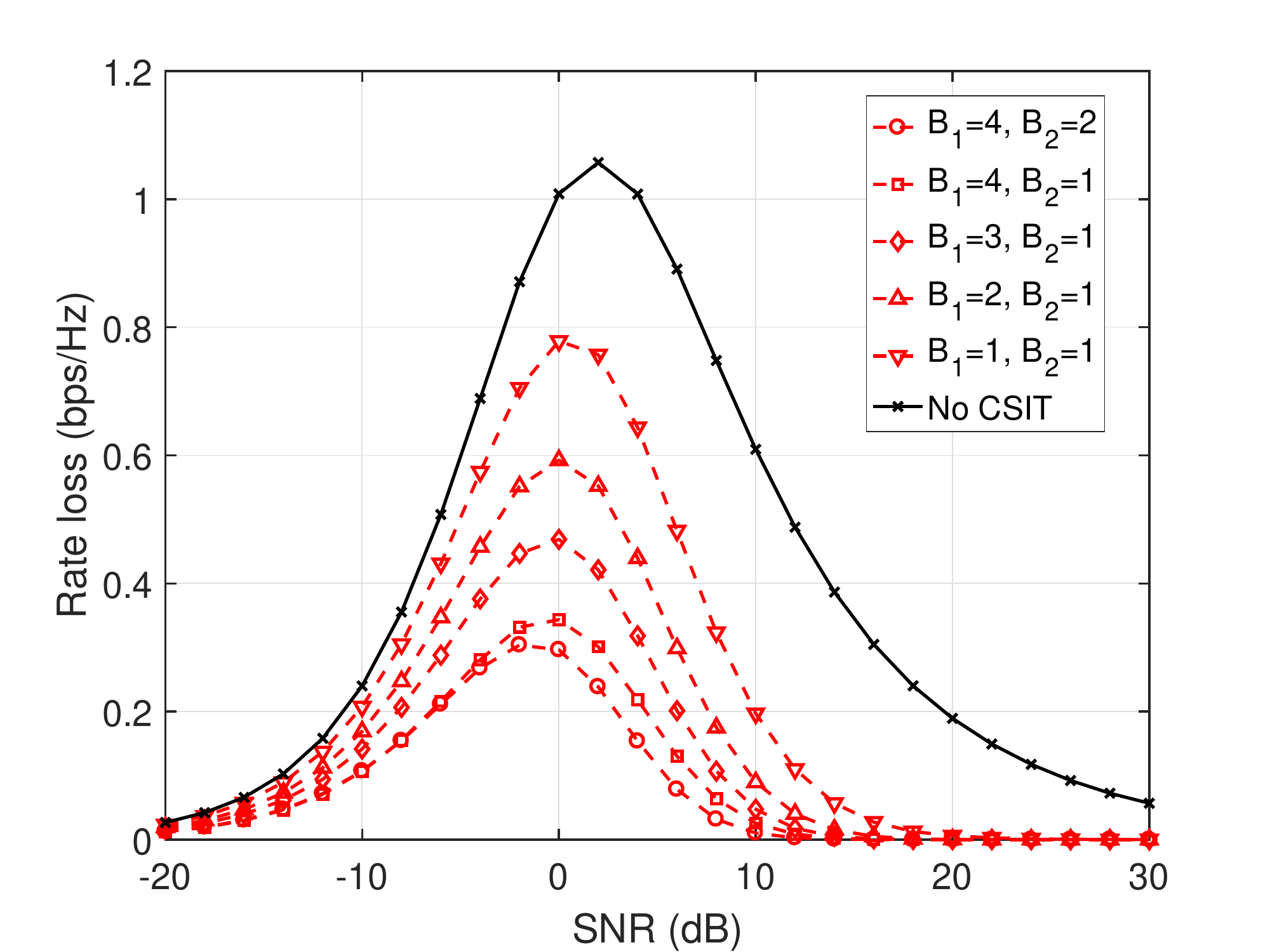}
		\vspace{-0.1cm}
		\centering
		\caption{The rate loss $\Delta R_{\mr{MISO}}(1,B)$ incurred by finite-rate feedback in a single-user MISO system when $\Nt=4$ and the ADC resolution is 1-bit. The rate loss is most severe at medium SNR.
			%The entries of channel $\bh$ follow IID circularly symmetric complex Gaussian distribution with unit variance.
		}\label{fig:Capacity_loss_Nt_4}
	\end{centering}
	\vspace{-0.3cm}
\end{figure}

%Another metric is bit error rate.
%The bit error rate is
%\begin{eqnarray}
%  Q\left(\sqrt{\gamma \left|h\right|^2} \right).
%\end{eqnarray}
%
%The bit error rate is
%\begin{eqnarray}
%  \frac{1}{2} Q\left(\sqrt{\gamma  \left|h\right|^2 \left(1 - \sin 2 \theta \right)}\right) + \frac{1}{2} Q\left(\sqrt{\gamma  \left|h\right|^2 \left(1 + \sin 2 \theta \right)}\right)
%\end{eqnarray}
%
%{\color{blue}{The average BER is... cannot find a close-form expression}}

\begin{figure}[t]
	\begin{centering}
		\includegraphics[width=0.7\columnwidth]{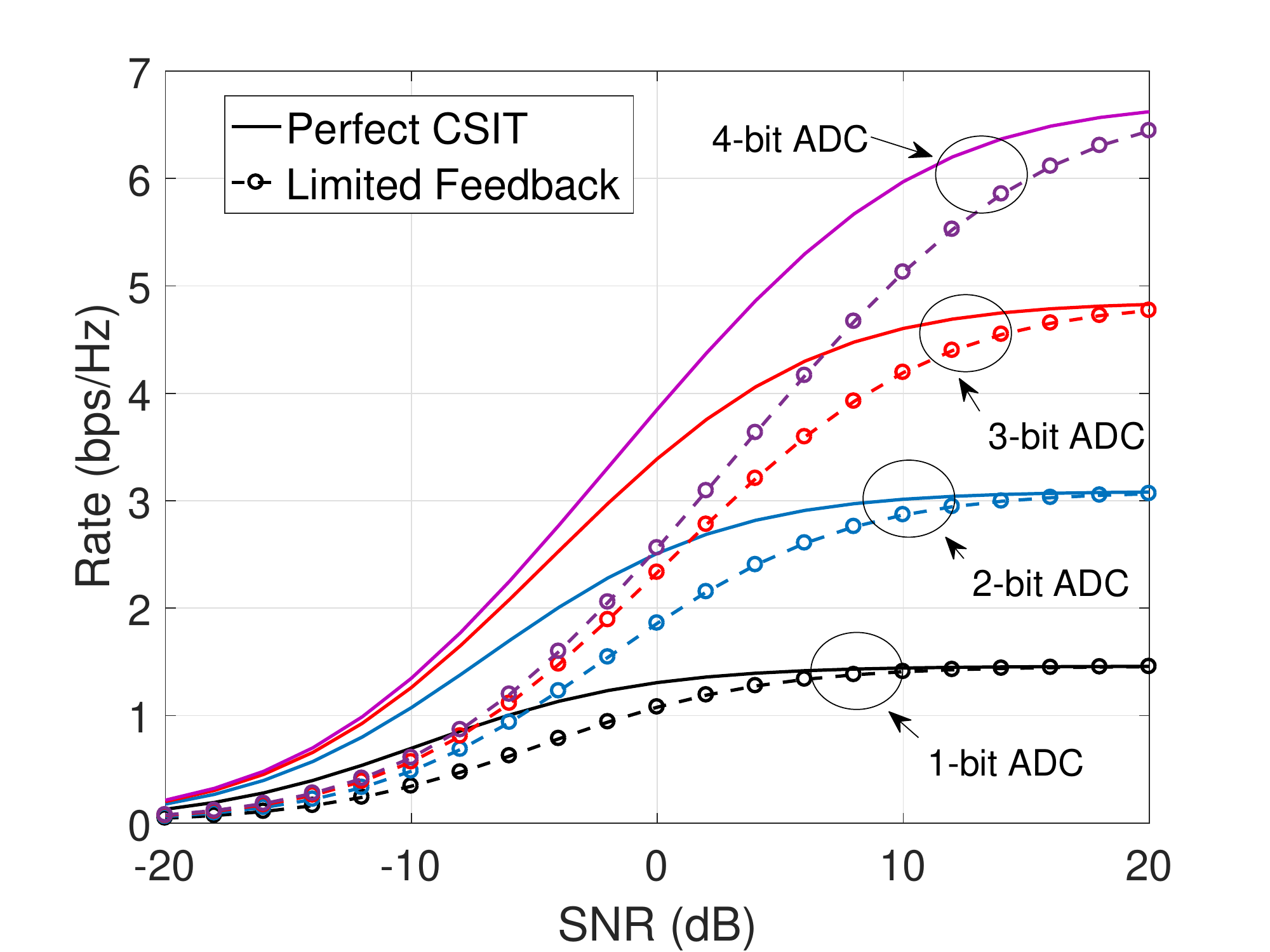}
		\vspace{-0.1cm}
		\centering
		\caption{The achievable rates of a single-user MISO system with CSIT and limited feedback when $\Nt=16$ and $B=8$. The ADC resolution increases from 1-bit to 4-bit.
			%The entries of channel $\bh$ follow IID circularly symmetric complex Gaussian distribution with unit variance.
		}\label{fig:SU_MISO_Rate_vs_SNR_Nt_16_B_8}
	\end{centering}
	\vspace{-0.3cm}
\end{figure}

\begin{figure}[t]
	\begin{centering}
		\includegraphics[width=0.7\columnwidth]{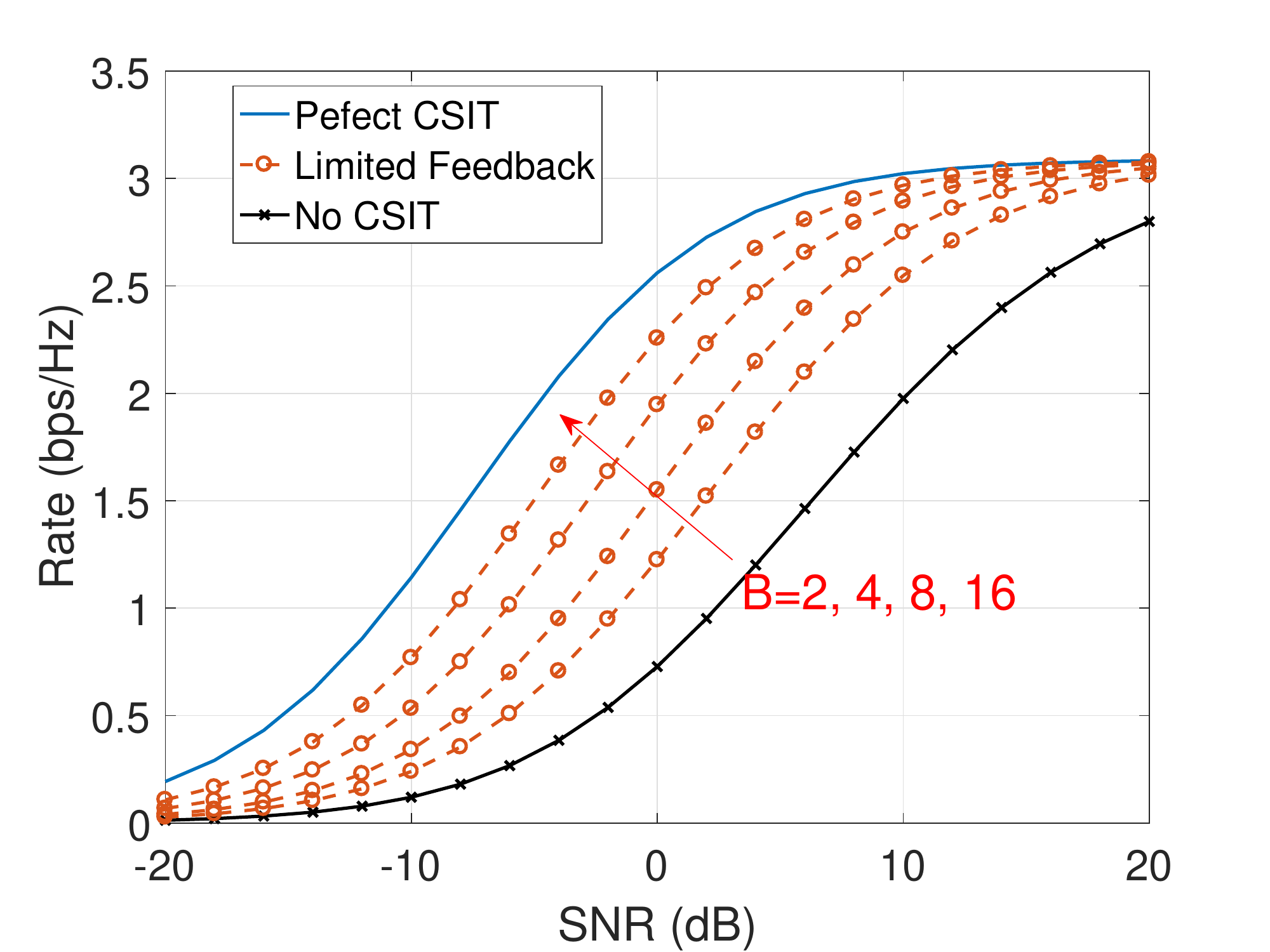}
		\vspace{-0.1cm}
		\centering
		\caption{The achievable rate of a single-user MISO system with CSIT, no CSIT and limited feedback when $\Nt=16$ and the ADC resolution is 2-bit. The curves corresponding to $B=0$ (No CSIT), $B=2$, $B=4$, $B=8$, $B=16$, $B=\infty$ (Perfect CSIT) are plotted.
		}\label{fig:SU_MISO_Rate_vs_SNR_Nt_16_b_2}
	\end{centering}
	\vspace{-0.3cm}
\end{figure}

%\begin{figure}[t]
%	\centering
%	\subfigure[$s$ is real-valued]{
%		\includegraphics[width=0.7\columnwidth]{Figures/Fig_SU_MIMO_Rate_vs_SNR_Nt_16_Nr_1_4_16_b_2_B_4_real_valued}
%		\label{fig:SU_MIMO_Rate_vs_SNR_Nt_16_Nr_1_4_16_b_2_B_4_real_valued}
%	}
%	\subfigure[$s$ is complex-valued]{
%		\includegraphics[width=0.7\columnwidth]{Figures/Fig_SU_MIMO_Rate_vs_SNR_Nt_16_Nr_1_4_16_b_2_B_4.eps}
%		\label{fig:SU_MIMO_Rate_vs_SNR_Nt_16_Nr_1_4_16_b_2_B_4}
%	}
%	\caption{The capacity of a point-to-point MIMO system with CSIT and limited feedback when $\Nt=16$, $b=2$ and $B=4$.}
%	\label{fig:SU_MIMO_Rate_vs_SNR}
%\end{figure}

\begin{figure}[t]
	\begin{centering}
		\includegraphics[width=0.7\columnwidth]{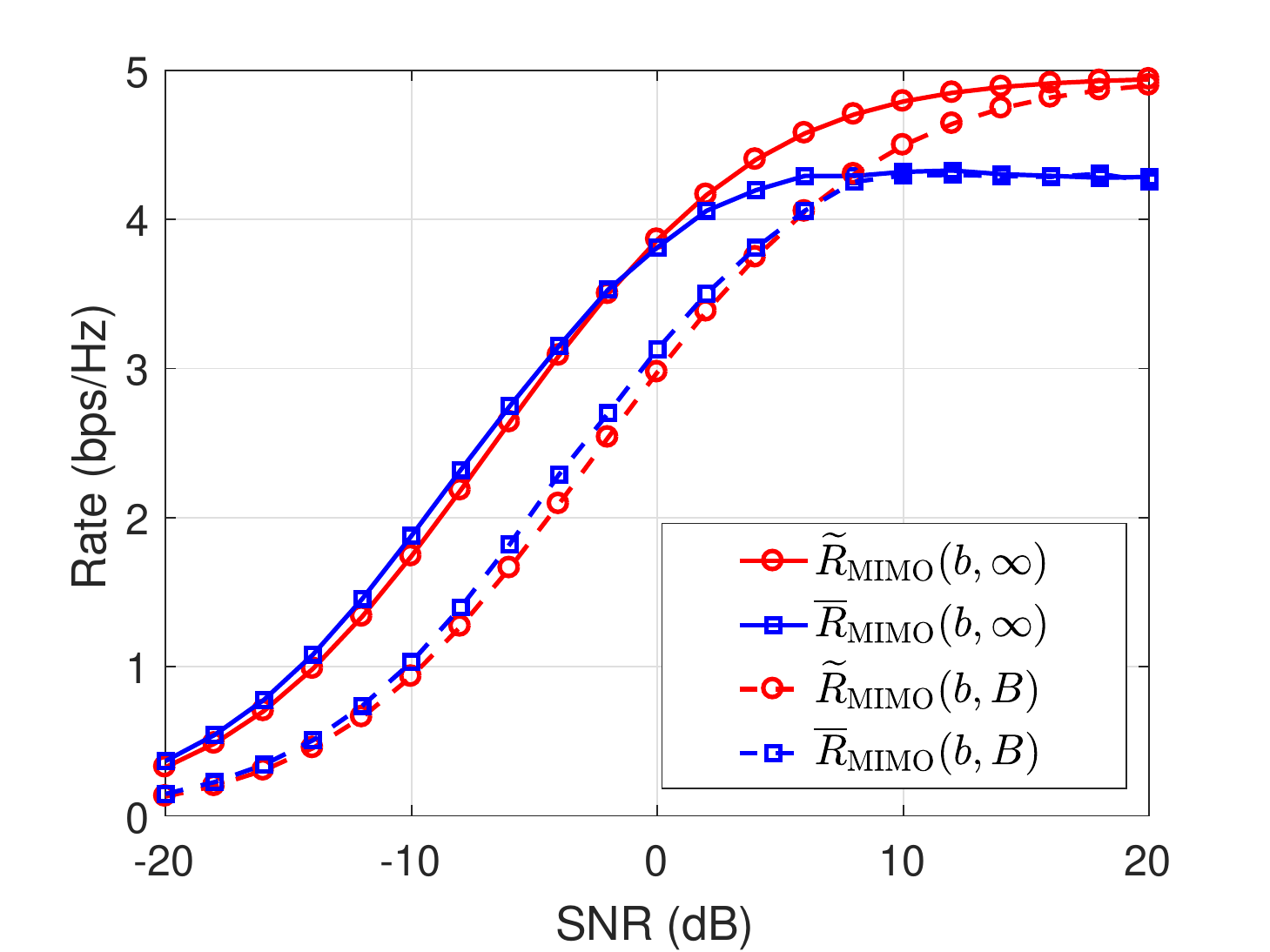}
		\vspace{-0.1cm}
		\centering
		\caption{The achievable rate of a single-user MIMO system with CSIT and limited feedback when $\Nt=16$, $\Nr=4$, $b=2$ and $B=4$. The lower bound $\widetilde{R}_{\mathrm{MIMO}}$ \eqref{eq:MIMO_Rate_lb} and approximate lower bound $\overline{R}_{\mathrm{MIMO}}$ \eqref{eq:MIMO_Rate_approx_lb} are plotted for comparison. They are close at low and medium SNR but have a gap at high SNR.
		}\label{fig:SU_MIMO_Rate_vs_SNR_Nt_16_Nr_4_b_2_B_4}
	\end{centering}
	\vspace{-0.3cm}
\end{figure}

\begin{figure}[t]
	\begin{centering}
		\includegraphics[width=0.7\columnwidth]{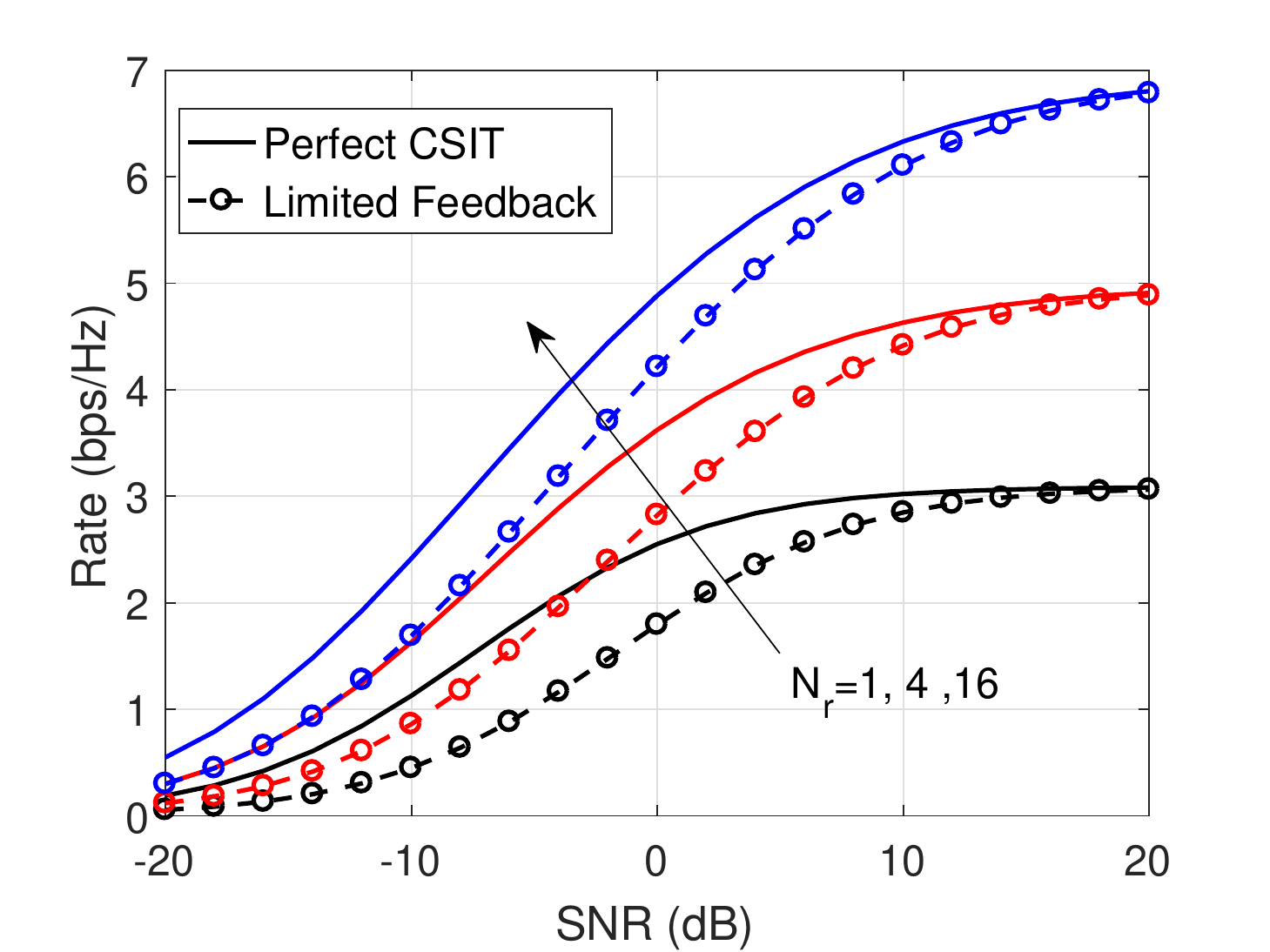}
		\vspace{-0.1cm}
		\centering
		\caption{The achievable rate of a single-user MIMO system with CSIT and limited feedback when $\Nt=16$, $b=2$ and $B=4$. Three cases where receiver has single antenna, four antennas and 16 antennas are shown.
		}\label{fig:SU_MIMO_Rate_vs_SNR_Nt_16_Nr_1_4_16_b_2_B_4}
	\end{centering}
	\vspace{-0.3cm}
\end{figure}

\begin{figure}[t]
	\begin{centering}
		\includegraphics[width=0.7\columnwidth]{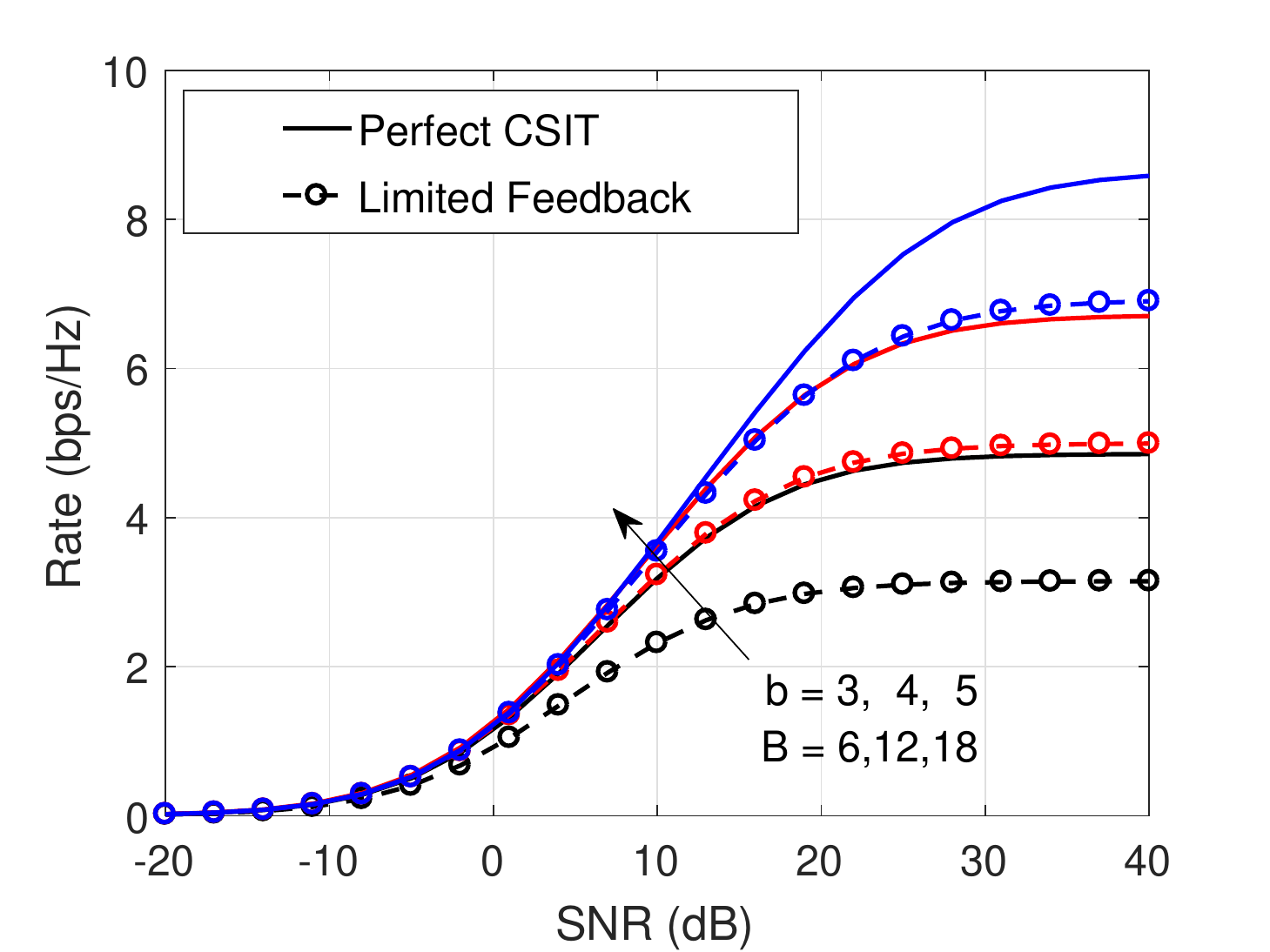}
		\vspace{-0.1cm}
		\centering
		\caption{The achievable rate of a multi-user MISO system with perfect CSIT and limited feedback when $\Nt=4$ and $K=2$. When there is perfect CSIT, the figure shows three cases where $b=3, 4, 5$. When there is limited feedback, the figure shows three cases where `$b=3, B=6$', `$b=4, B=12$' and `$b=5, B=18$' satisfying $B=6b-12$.
		}\label{fig:MU_MISO_Rate_vs_SNR_Nt_4_K_2}
	\end{centering}
	\vspace{-1cm}
\end{figure}

\section{Simulation Results}

In this section, we evaluate the performance of the proposed limited feedback methods for different configurations. We compute the achievable rate for each channel realization then averaged over 1000 channel realizations assuming Rayleigh fading, i.e., elements of $\bh$ and $\bH$ follow IID Gaussian distribution $\mathcal{CN}(0, 1)$. In the figures, $\mathrm{SNR (dB)} \triangleq 10 \log_{10} \frac{\Pt}{\sigma_n^2}$.

In Fig. \ref{fig:SISO_Capacity}, we  compare the capacities of 1-bit quantized SISO channel with perfect \ac{CSIT}, limited feedback and no CSIT.  As shown in the figure, the rate with two bits feedback is almost same as that with perfect CSIT. In addition, even with a single bit of feedback, the power loss is very small, i.e., less than 1 $\mr{dB}$. Without CSIT, the rate loss is much larger, especially at high SNR. Taking into account both the feedback overhead and the rate loss, it is reasonable to set $B=1$ in practice.

Figs. \ref{fig:MISO_Capacity_1bit_ADC} - \ref{fig:Capacity_loss_Nt_4} show the performance of the proposed limited feedback scheme in 1-bit quantized MISO channel. 
In \figref{fig:MISO_Capacity_Nt_4}, we plot the capacities of MISO with four antennas and four bits feedback. Five different allocations of the feedback bits are compared. It is found that the case `$B_1=3, B_2=1$' has the best performance with power loss around $3$ $\mr{dB}$, which is consistent with our analysis in \eqref{eq:MISO_power_loss} which states the power loss is upper bounded by $6$ $\mr{dB}$. In Fig. \ref{fig:MISO_Capacity_Nt_16}, we show another example with 16 antennas and 16 bits feedback. It is shown that the case `$B_1=15, B_2=1$' is the best one. We therefore conclude that more bits should be assigned to feed back the channel direction information. If there is no bit to feed back the residual phase information, i.e., `$B_1=\Nt, B_2=0$', the achievable rate is much lower than that of `$B_1 = \Nt-1, B_2=1$' in the medium and high SNR regimes. This implies that the phase information is important at the medium and high SNR regimes and at least one bit should be assigned to feed back this information.

In \figref{fig:Capacity_loss_Nt_4}, we show the rate loss for different values of $B_1$ and $B_2$. As expected, the rate loss decreases as $B_1$ and $B_2$ increase. We also find that at high SNR, the rate losses incurred by limited feedback converge to zero. For instance, when the transmitter power is larger than $11$ $\mr{dB}$, even with $B_1=B_2=1$, the rate loss is less than $0.2$ $\mr{bps/Hz}$, which implies that the rate with only two bit feedback achieves $90\%$ of the rate with perfect CSIT. The result verifies our analysis in \eqref{eq:MISO_high_SNR_fb_example}.
Note that the rate loss at low SNR is also small since the channel capacity is small.

In \figref{fig:SU_MISO_Rate_vs_SNR_Nt_16_B_8}, we show the average achievable rates with perfect CSIT and limited feedback in a single-user MISO channel. First, the rate with perfect CSIT converges to $\log_2 \left( \frac{1}{\eta_b}\right)$, which is $1.46$, $3.09$, $4.86$, $6.72$ bps/Hz when $b=1, 2, 3, 4$. Note that these values are less than the theoretical upper bound $2b$ bps/Hz because Gaussian signaling is suboptimal.
Second, at high SNR (for instance, 10 dB when $b=1$, 20 dB when $b=4$), there is almost no rate loss between the perfect CSIT and limited feedback cases since the quantization noise dominates the AWGN noise in this regime. Third, in the low SNR regime ($<0$ dB), we see there is a constant horizontal distance between each pair of solid curve and dashed curve which implies that there is a constant power loss incurred by limited feedback. This is because the AWGN noise dominates the performance and the results from previous work assuming infinite-bit ADCs \cite{Au-Yeung_TWC07, Mukkavilli_IT03} then apply.
\figref{fig:SU_MISO_Rate_vs_SNR_Nt_16_b_2} shows the achievable rate of a single-user MISO with 2-bit ADCs for different number of feedback bits. As $B$ increases from 2 to 16, we see that the SNR loss decrease from around $10$ dB to $3$ dB in the medium SNR regime. These numbers are close to $10 \log_{10} \left(1-2^{-\frac{B}{\Nt-1}}\right)$ dB given in our analyses.

\figref{fig:SU_MIMO_Rate_vs_SNR_Nt_16_Nr_4_b_2_B_4} shows the achievable rate of a MIMO with perfect CSIT and limited feedback. First, the lower bound in \eqref{eq:MIMO_Rate_lb} and the approximate lower bound in \eqref{eq:MIMO_Rate_approx_lb} are compared. It is seen that when the SNR is less than $0$ dB, the approximate is quite accurate. Second, at high SNR (20 dB), the two bounds both saturate and there is no loss due to limited feedback.
\figref{fig:SU_MIMO_Rate_vs_SNR_Nt_16_Nr_1_4_16_b_2_B_4} presents the achievable rate of a MIMO channel with 2-bit ADCs for different number of receive antennas. First, the rate loss is not significant at high SNR which is consistent with our analyses. Second, we find that the rate saturates to $\log_2 \left(1 +　\frac{1-\eta_b}{\eta_b} \Nr\right) \approx \log_2 \left(1 +　7.51 \Nr\right)$ bps/Hz, which is $3.09$, $4.96$ and $6.92$ bps/Hz when $\Nr=1, 4, 16$.

In \figref{fig:MU_MISO_Rate_vs_SNR_Nt_4_K_2}, we show the achievable rates in a multi-user MISO channel. The number feedback bits is chosen as $B = 2(\Nt-1)b - 12 = 6b-12$. First, apart from the single-user case, there is a gap at high SNR between the case of perfect CSIT and limited feedback due to the inter-user interference. Second, the gaps between each pair of curves are all around $1.7$ bps/Hz, which verifies our analytical result in \eqref{eq:MU_MISO_scaling_law} stating that $B$ should increase $2(\Nt-1)$ bits if $b$ increases one bit. Third, at low SNR ($<0$ dB), the power loss is small for three cases, which validates our results in \eqref{MU_MISO_Rate_loss_low_SNR} saying that the power loss is $10 \log_{10} \left(1 - 2^{-\frac{B}{\Nt-1}}\right)$ dB, which is around $-1.25$ dB when $B=6$, $-0.28$ dB when $B=12$, and $-0.07$ dB when $B=18$.

%%%%%%%%%%%%%%%%%%%%%%%%%%%%%%%%%%%%
\section{Conclusions}
%%%%%%%%%%%%%%%%%%%%%%%%%%%%%%%%%%%%
In this paper, we developed and analyzed limited feedback methods in multiple-antenna system with few-bit ADCs. First, we proposed an approach for limited feedback in SISO and MISO channels with 1-bit ADCs. For the SISO channel, only the phase of the channel is quantized while in the MISO channel, the channel direction and residual phase are both quantized and fed back to the transmitter. This design, however, cannot be extended to the channel with more than 1-bit ADCs because the optimal signaling in this case is unknown. Therefore, by assuming the transmitted signal has a Gaussian distribution and the quantization noise is worst-case Gaussian distributed, a lower bound of the capacity was derived. The receiver only feedbacks the channel direction and the limited feedback loss was analyzed. 
%
%We evaluated the power and capacity losses incurred by the use of limited feedback. Based on our analyses and simulation results, we made two important observations. First, feeding back only one bit for the phase (or residual phase in MISO channel) is enough to guarantee good performance in the examples considered. Second, when the capacity of the quantized channel is saturated, the required number of feedback bits guaranteeing a small capacity loss decreases with SNR.
%

Second, we evaluated the achievable rate in single-user MIMO channel with limited feedback. We provided a method to compute the covariance matrix of the quantization output signal if the input signal is circularly symmetric. Then a heuristic feedback method, where the leading eigenvector was quantized, was proposed.
Last, we analyzed the achievable rate in multi-user MISO systems with multi-bit ADC and limited feedback. The results are similar to those with infinite-bit ADC at low SNR except for an additional power loss $10 \log_{10} \left(1-\eta_b\right)$ dB incurred by low resolution ADCs. At high SNR, however, the quantization noise dominates and therefore the results are very different from the case with infinite-bit ADCs. 
%
%In the single-user channel, we found that the achievable rate saturates to a upper bound determined by signal-to-quantization ratio of the ADCs. In the multi-user case, we found that that the number of bits per feedback should increase linearly with the ADC resolution to limit the rate loss at high SNR.

We obtained several design guidelines based on our analyses and simulation results. %For example, the scaling law of the number of feedback bits versus the ADC resolution to restrict the performance loss. 
The first insight is that the rate loss is most severe in the medium SNR regime and more feedback bits are needed in that regime. The second insight is that for the multi-user MISO channel, the number of feedback bits should increase linearly with ADC resolution to limit the rate loss based on the scaling law derived in this paper.

There are several potential directions for future work. Our numerical results were based on the IID Gaussian channel with small numbers of antennas. In mmWave systems - a promising application of 1-bit ADCs - the channels will likely be correlated depending on the number of scattering clusters and the angle spread. It would be interesting to develop techniques that also work for large correlated channels. It is also interesting to extend our result from the narrowband channel to the broadband channel.
%A natural extension of our work would be to MIMO communication channels. This is complicated by the complicated structure of the capacity-optimum signaling distribution and the potential for different choices of precoders (see \cite[Section IV]{Mo_Jianhua_TSP15}).
Another possible direction is to combine the two separate stages, channel estimation and limited feedback, together. In this case, the feedback bits are decided directly by the ADC outputs instead of the estimated CSI at the receiver.

\bibliographystyle{IEEEtran}
\bibliography{IEEEabrv,One_bit_Quantization}
\end{document}